\documentclass[12pt]{article}                     
\usepackage{graphicx}

\newcommand{\nr}{\hspace*{0.7 true cm}}    

\newcommand{\Ma}{{\cal M}}

\newcommand{\0}{\phantom{0}}                      

\newcommand{\ADN} {{\it At. Nucl. Data Tables }}
\newcommand{\AIP} {{\it AIP Conf. Proc. }}
\newcommand{\APH} {{\it Acta Phys. Hungarica }}
\newcommand{\EPJA}{{\it Eur. Phys. Journal A }}
\newcommand{\EPJC}{{\it Eur. Phys. Journal C }}
\newcommand{\IOP} {{\it Inst. Phys. Conf. Series }}
\newcommand{\NIM} {{\it Nucl. Instrum. Meth. }}
\newcommand{\NIMA}{{\it Nucl. Instrum. Meth. A }}
\newcommand{\NIMB}{{\it Nucl. Instrum. Meth. B }}
\newcommand{\NPA} {{\it Nucl. Phys. A }}
\newcommand{\PLA} {{\it Phys. Lett. A }}
\newcommand{\PLB} {{\it Phys. Lett. B }}
\newcommand{\PR}  {{\it Phys. Rev.}}
\newcommand{\PRC} {{\it Phys. Rev. C }}
\newcommand{\PRL} {{\it Phys. Rev. Lett. }}
\newcommand{\PS}  {{\it Phys. Scr. }}
\newcommand{\RMP} {{\it Rev. Mod. Phys. }}
\newcommand{\RRP} {{\it Rev. Roum. Phys. }}
\newcommand{\RSI} {{\it Rev. Sci. Instr. }}
\newcommand{\ZP}  {{\it Z. Phys. }}

\begin{document}
   \begin{center}
   ~ \vspace {3mm}
   {\Large \bf
       A Lecture on the Evaluation of Atomic Masses}    \vspace {2mm}

   {\large Georges Audi}                   \vspace {2mm}

   {\small First edition: September 2000, updated December 2, 2004} \vspace {2mm}

   {\small\it
       Centre de Spectrom\'etrie Nucl\'eaire et de
       Spectrom\'etrie de Masse, CSNSM, \\ IN2P3-CNRS, et UPS,
       B\^atiment 108, F-91405 Orsay Campus, France}  \vspace {5mm}

   \end{center}

   \begin{abstract}
   \noindent
   The ensemble of experimental data on the 2830 nuclides which have
been observed since the beginning of Nuclear Physics are being
evaluated, according to their nature, by different methods and by
different groups.
   The two ``horizontal'' evaluations in which I am involved: the Atomic
Mass Evaluation {\sc Ame} and the {\sc Nubase} evaluation belong to the
class of ``static" nuclear data.
   In this lecture I will explain and discuss in detail
the philosophy, the strategies and the procedures used in the evaluation
of atomic masses.
   \\\\
   {\bf R\'esum\'e}
   {\it L'\'evaluation des masses atomiques} -
   Les donn\'ees exp\'erimentales sur les 2830 nucl\'eides observ\'es
depuis les d\'ebuts de la Physique Nucl\'eaire sont \'evalu\'ees,
suivant leur nature, par diff\'erentes m\'ethodes et par diff\'erents
groupes.
   Les deux \'evaluations ``horizontales'' dans les- quelles je suis
impliqu\'e : l'\'Evaluation des Masses Atomiques {\sc Ame} et
l'\'evaluation {\sc Nubase} appartiennent \`a la classe des donn\'ees
nucl\'eaires ``statiques".
   Dans ce cours je vais expliquer et discuter de mani\`ere approfondie
la philosophie, les strat\'egies et les proc\'edures utilis\'ees dans
l'\'evaluation des masses atomiques.  \\

   \noindent
   {\bf PACS.} ~ 21.10.Dr ; 21.10.Hw ; 21.10.Tg ; 23.40.-s ; 23.50.+z ; 23.60.+e \\
   {\em Keywords:} ~ Binding energies - atomic masses - horizontal evaluation -
                     nuclear~data~- least-squares~- predictions of unknown masses
   \end{abstract}

   \section{Introduction}

   Nuclear Physics started a little bit more than 100 years ago with the
discoveries of natural radioactivity by Henri Becquerel and Pierre and
Marie Curie; and of the atomic nucleus in 1911 by Ernest Rutherford with
his assistants Ernest Marsden and Hans Geiger.
   First, it was a science of curiosity exhibiting phenomena unusual for
that time.
   It is not until the late thirties, well after the discovery of
artificial radioactivity by Fr\'ed\'eric and Ir\`ene Joliot-Curie, that
the research in that domain tended to accelerate drastically and that
Nuclear Physics became more and more a quantitative science.

\begin{figure}[htb]   
   \begin{center}
   \includegraphics[width=15 cm]{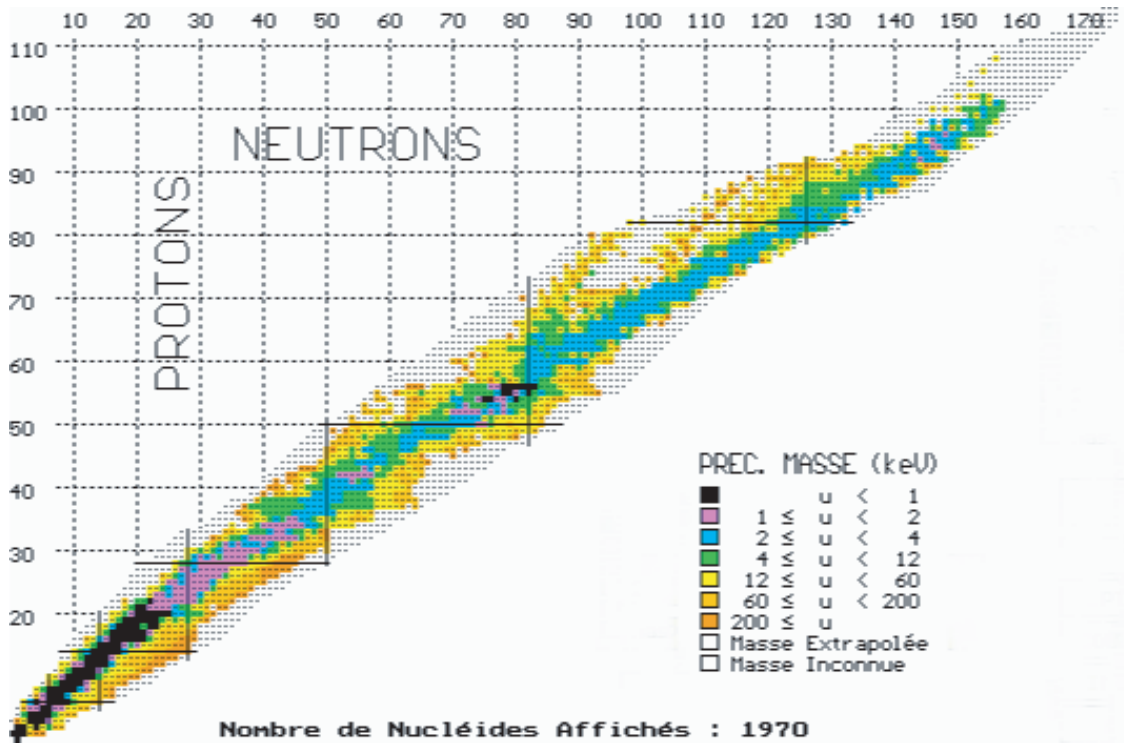}
   \caption[]{\footnotesize
   Chart of nuclides for the precision on masses.
   Only the known masses are colored, exhibiting crudely the narrowness
of the valley of our knowledge in this immense landscape.
   Would these 1970 known masses been scattered around in the ($N,Z$)
plane, our understanding of the nucleus would have been completely
changed.
   }\label{fig:Mass}
   \end{center}
   \end{figure}

   Since then, scientists have accumulated a huge amount of data on a
large number of nuclides.
   Today there are some 2830 variations on the combination of
protons and neutrons that have been observed.
   Although this number seems large, specially compared to the 6\,000 to
7\,000 that are predicted to exist, one should be aware that the numbers
of protons and neutrons constituting a nuclide are not really independent.
   Their special correlation form a relatively narrow band around a line
called the bottom of the valley of stability.
   In Fig.\,\ref{fig:Mass} this is illustrated for the known masses
(colored ones) across the chart of nuclides.
   In other words, nuclear data put almost no constraint in isospin on
nuclear models.
   From there follows the tendency of nuclear physicists to study
nuclides at some distance from that line, which are called {\em exotic}
nuclides.

   Sometimes remeasurement of the same physical quantity improved a
previous result; sometimes it entered in conflict with it.
   The interest of the physicist has also evolved with time: the
quantities considered varied importantly, scanning all sort of data from
cross sections to masses, from half-lives to magnetic moments, from
radii to superdeformed bands.

   Thus, we are left nowadays with an enormous quantity of information
on the atomic nucleus that need to be sorted, treated in a homogeneous
way, while keeping traceability of the conditions under which they were
obtained.
   When necessary, different data yielding values for the same physical
quantity need to be compared, combined or averaged to derive an
adopted value.
   Such values will be used in domains of physics that can be very far
from nuclear physics, like half-lives in geo-chronology, cross-sections
in proton-therapy, or masses in the determination of the $\alpha$ fine
structure constant.

   There are two classes of nuclear data: one class is for data related
to nuclides at rest (or almost at rest); and the other class is for
those related to nuclidic dynamics.
   In the first class, one finds ground-state and level properties,
whereas the second encompasses reaction properties and mechanisms.

   Nuclear ground-state masses and radii;
   magnetic moments;
   thermal neutron capture cross-sections;
   half-lives, spins and parities of excited and ground-state levels;
   the relative position (excitation energies) of these levels;
   their decay modes and the relative intensities of these decays;
   the transition probabilities from one level to another and the level width;
   the deformations;
   all fall in the category of what could be called the ``static''
nuclear properties.

   Total and differential (in energy and in angle) reaction cross-sections;
   reaction mechanisms; and
   spectroscopic factors
   could be grouped in the class of ``dynamic'' nuclear properties.

   Certainly, one single experiment, for example a nuclear reaction
study, can yield data for both `static' and `dynamic' properties.

   It is out of the scope of the present lecture to cover all aspects of
nuclear properties and nuclear data.
   The fine structure of ``static'' nuclear data will be shortly
described and the authors of the various evaluations presented.
   Then I will center this lecture on the two ``horizontal'' evaluations
in which I am involved: the atomic mass evaluation {\sc Ame} and the
{\sc Nubase} evaluation, both being strongly related, particularly when
considering isomers.

   \section{``Static'' nuclear data}

   \subsection{The {\sc Nsdd}: data for nuclear structure}

\begin{figure}[htb]   
   \begin{center}
   \includegraphics[width=10 cm]{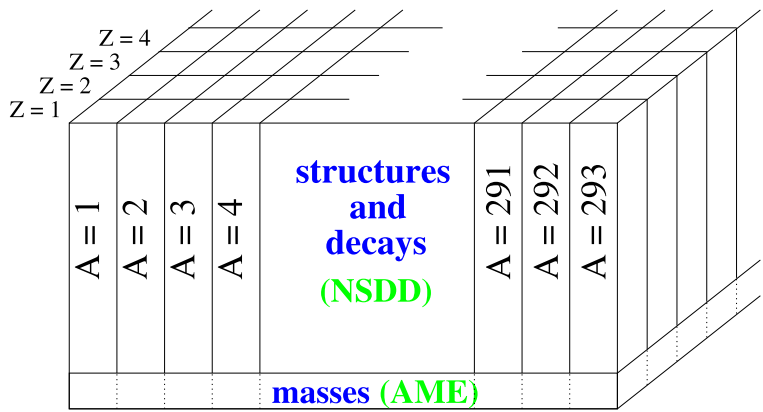}
   \caption[]{\footnotesize
   Schematic representation of all the available ``static" nuclear data
(structure, decay, mass, radius, moments,\ldots).
   Each nuclide is represented as a building with its ground-state at
the ground floor.
   The mass evaluation is represented on the ground floor, across all
buildings.
   It includes also data for upper levels if they represent an energy
relation to another nuclide, like a foot-bridge between two buildings
that will allow to derive the level difference between their ground
floors.
   }\label{fig:static}
   \end{center}
   \end{figure}

   The amount of data to be considered for nuclear structure is huge.
   They are represented schematically in Fig.\,\ref{fig:static} for each
nuclide as one column containing all levels from the ground-state at the
bottom of that column to the highest known excited state.
   All the known properties for each of the levels are included.
   Very early, it was found convenient to organize their evaluation in a
network, splitting these data according to the mass of the nuclides, the
$A$-chains.
   Such a division makes sense, since most of the decay relations among
nuclides are $\beta$-decays where $A$ is conserved.
   This is, of course, less true for heavier nuclides where
$\alpha$-decay is the dominant decay-mode connecting an $A$-nuclide to
an $A-4$ daughter.
   This structure is the one adopted by the Nuclear Structure and Decay
Data network (the {\sc Nsdd}) organized internationally under the
auspices of the {\sc Iaea} in Vienna.
   An $A$-chain or a group of successive $A$-chains is put under the
responsibility of one member of the network.
   His or her evaluation is refereed by another member of the network
before publication in the journal `Nuclear Data Sheets' (or in the
`Nuclear Physics' journal for $A~\le~44$).
   At the same time the computer files of the evaluation (the
{\sc Ensdf}: `Evaluated Nuclear Structure Data Files') are made
available at the {\sc Nndc}-Brookhaven \cite{90Burrows}.
   In this evaluation network, most of the ``static'' nuclear data are
being considered.

   The {\sc Nsdd} evaluation for each nuclide is not dependent, at first
order, on the properties of a neighboring nuclide, except when there is
a decay relation with that neighbor.
   Such evaluation, conducted nuclide by nuclide, is called a `vertical'
evaluation.

   \subsection{The atomic mass evaluation {\sc Ame}} \label{sect:Ame-gen}

   However, the evaluation of data for energy relations
between nuclides is more complex due to numerous links that
overdetermine the system and exhibit sometimes inconsistencies
among data.
   This ensemble of energy relations is taken into account in the
`horizontal' structure of the Atomic Mass Evaluation {\sc Ame}
\cite{Ame03,Ame03a}.
   By `horizontal' one means that a unique nuclear property is being
considered across the whole chart of nuclides, here the ground-state
masses.
   Only such a structure allows to encompass all types of connections
among nuclides, whether derived from $\beta$-decays, $\alpha$-decays,
thermal neutron-capture, reaction energies, or mass-spectrometry where
any nuclide, e.g. $^{200}$Hg can be connected to a molecule like
$^{12}$C$^{13}$C$^{35}$Cl$_5$ or, in a Penning trap mass spectrometer,
to $^{208}$Pb.

   The {\sc Ame}, the main subject of this lecture, will be developed
below, in Section ~\ref{sect:Ame}.

   \subsection{The isomers in the {\sc Ame} and the emergence of {\sc Nubase}}  \label{sect:Nub}

   At the interface between the {\sc Nsdd} and the {\sc Ame}, one is
faced with the problem of identifying - in some difficult cases - which
state is the ground-state.
   The isomer matter is a continuous subject of worry in the {\sc Ame},
since a mistreatment can have important consequences on the ground-state
masses.
   Where isomers occur, one has to be careful to
check which one is involved in reported experimental data, such as
$\alpha$- and $\beta$-decay energies.
   Cases have occurred where authors were not (yet) aware of isomeric
complications.
   The matter of isomerism became even more important, when mass
spectrometric methods were developed to measure masses of exotic atoms
far from $\beta$-stability and therefore having small half-lives.
   The resolution in the spectrometers is limited, and often
insufficient to separate isomers.
   Then, one so obtains an average mass for the isomeric pair.
   A mass of the ground-state, our primary purpose, can then only be
derived if one has information on the excitation energy and on the
production rates of the isomers.
   And in cases where e.g. the excitation energy was not known, it may
be estimated, see below.

   When an isomer decays by an internal transition, there is no
ambiguity and the assignment as well as the excitation energy is given
by the {\sc Nsdd} evaluators.
   However, when a connection to the ground-state cannot be obtained,
most often a decay energy to (and sometimes from) a different nuclide
can be measured (generally with less precision).
   In the latter case one enters the domain of the {\sc Ame}, where
combination of the energy relations of the two long-lived levels to
their daughters (or to their parents) with the masses of the latter,
allows to derive the masses of both states, thus an excitation energy
(and, in general, an ordering).

   Up to the 1993 mass table, the {\sc Ame} was not concerned with all
known cases of isomerism, but only in those that were relevant to the
determination of the ground-state masses.
   In 1992 it was decided, after discussion with the {\sc Nsdd}
evaluators, to include all isomers for which the excitation energy ``is
not derived from $\gamma$-transition energy measurements ($\gamma$-rays
and conversion electron transitions), and also those for which the
precision in $\gamma$-transitions is not decidedly better than that of
particle decay or reaction energies leading to them" \cite{Ame95}.

   However, differences in isomer assignment between the {\sc Nsdd} and
the {\sc Ame} evaluations cannot be all removed at once, since the
renewal of all $A$-chains in {\sc Nsdd} can take several years.
   In the meantime also, new experiments can yield information that
could change some assignments.
   Here a `horizontal' evaluation should help.

   The isomer matter was one of the main reasons for setting up in 1993
the {\sc Nubase} collaboration leading to a thorough examination and
evaluation of those ground-state and isomeric properties that can help
in identifying which state is the ground-state and which states are
involved in a mass measurement \cite{Nubase03}.
   {\sc Nubase} appears thus as a `horizontal' database for several
nuclear properties: masses, excitation energies of isomers, half-lives,
spins and parities, decay modes and their intensities.
   Applications extend from the {\sc Ame} to nuclear reactors, waste
management, astrophysical nucleo-synthesis, and to preparation of
nuclear physics experiments.

   Setting up {\sc Nubase} allowed in several cases to predict the
existence of an unknown ground-state from trends of isomers in
neighboring nuclides, whereas only one long-lived state was reported.
   A typical example is $^{161}$Re, for which {\sc Nubase'97}
\cite{Nubase97} predicted a (${1/2}^+\#$) proton emitting state below an
observed 14~ms $\alpha$-decaying high-spin state.
   (Everywhere in {\sc Ame} and {\sc Nubase} the symbol \# is used to
flag values estimated from trends in systematics.)
   Since then, the 370 $\mu$s, ${1/2}^+$, proton emitting state was
reported with a mass 124~keV below the 14~ms state.
   For the latter a spin ${11/2}^-$ was also assigned \cite{97Irvine}.
   Similarly, the ${11/2}^-$ bandhead level discovered in
$^{127}$Pr \cite{98Morek} is almost certainly an excited isomer.
   We estimate for this isomer, from systematical trends, an excitation
energy of 600(200)\#~keV and a half-life of approximatively 50\#~ms.

   In some cases the value determined by the {\sc Ame} for the isomeric
excitation energy allows no decision as to which of the two isomers is
the ground-state.
   This is particularly the case when the uncertainty on the excitation
energy is large compared to that energy, e.g.:
     $E^m(^{82}$As)$  =  250 \pm 200 $~keV;
     $E^m(^{134}$Sb)$ =   80 \pm 110 $~keV;
     $E^m(^{154}$Pm)$ =  120 \pm 120 $~keV.

\begin{figure}[htb]   
   \begin{center}
   \includegraphics[height=4.3 cm]{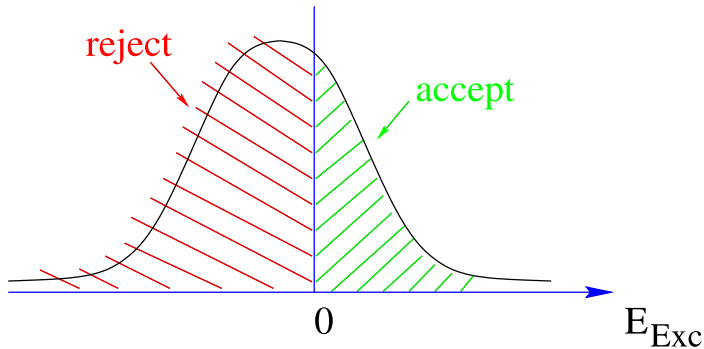}
   \caption[]{\footnotesize
   Truncated distribution of probability when there is a strong
indication about ordering of ground-state and isomer.
   }\label{fig:trunc}
   \end{center}
   \end{figure}

   Three main cases may occur.
   In the first case, there is no indication from the trends in
$J^{\pi}$ systematics of neighboring nuclides with same parities in $N$
and $Z$, and no preference for ground-state or excited state can be derived
from nuclear structure data.
   Then the adopted ordering, as a general rule, is such that the obtained
value for $E^m$ is positive.
   In the three examples above, $^{82}$As will then have its (5$^-$)
state located at 250$\pm$200~keV above the (1$^+$); in $^{134}$Sb the
(7$^-$) will be 80$\pm$110~keV above (0$^-$); and $^{154}$Pm's spin
(3,4) isomer 120$\pm$120~keV above the (0,1) ground-state.
   In the second case, one level could be preferred as ground-state from
consideration of the  trends  of systematics in  $J^{\pi}$.
   Then, the {\sc Nubase} evaluators accept the ordering given by these
trends, even if it may yield a (slightly) negative value for the
excitation energy, like in $^{108}$Rh (high spin state at
$-60\pm$110~keV).
   Such trends in systematics are still more useful for odd-$A$
nuclides, for which isomeric excitation energies of isotopes (if $N$ is
even) or, similarly, isotones follow usually a systematic course.
   This allows to derive estimates both for the relative position and
for the excitation energies where they are not known.
   Finally, there are cases where data exist on the order of the
isomers, e.g. if one of them is known to decay into the other one, or if
the Gallagher-Moszkowski rule \cite{58Gallagher} for relative positions of combinations
points strongly to one of the two as being the ground-state.
   Then the negative part, if any, of the distribution of probability
has to be rejected (Fig.\,\ref{fig:trunc}).
   Value and error are then calculated from the moments of the positive
part of the distribution.

   \subsection{Other `horizontal' evaluations}

   There might be other reasons for `horizontal' evaluations.
   The splitting of data among a large number of evaluators - like in
the {\sc Nsdd} network described above - does not always allow having a
completely consistent treatment of a given nuclear property through the
chart of nuclides.
   In addition, some quantities may fall at the border of the main
interest of such a network.
   This is the reason why a few `horizontal' compilations or evaluations
have been conducted for the benefit of the whole community.
   For example, one can quote the work of Otten 
\cite{89Otten} for isotope shift and hyperfine structure of spectral
lines and the deduced radii, spins and moments of nuclides in their
ground-state and long-lived isomeric states.
   An evaluation of isotope shifts has been published also by Aufmuth
and coworkers \cite{87Aufmuth}, and Raghavan \cite{89Raghavan} gave a
table of nuclear moments, updated recently by Stone \cite{99Stone}.
   More recent tables for nuclidic radii were published by Angeli
\cite{91Angeli} in 1991 and by Nadjakov {\it et al} \cite{94Nadjakov} in
1994.
   Two other `horizontal' evaluations are worth mentioning.
   One is the evaluation of isotopic abundances, by Holden
\cite{95Holden}.
   The second one is the evaluation of Raman and coworkers
\cite{89Raman} for the energy $E_{2^+}$ and the reduced electric
quadrupole transition probability $B(E2)$ of the first excited $2^+$
state in even-even nuclides.

   \section{The evaluation of atomic masses ({\sc Ame})} \label{sect:Ame}

   The atomic mass evaluation is particular when compared to the other
evaluations of data reviewed above, in that there are almost no absolute
determinations of masses.
   All mass determinations are relative measurements.
   Each experimental datum sets a relation in energy or mass among two
(in a few cases, more) nuclides.
   It can be therefore represented by one link among these two nuclides.
   The ensemble of these links generates a highly entangled network.
   This is the reason why, as I mentioned earlier
(cf.~Section~\ref{sect:Ame-gen}), a `horizontal' evaluation is essential.

   I will not enter in details in the different types of mass
experiments, since there are other lectures devoted to this
subject \cite{Orr-Lepine}.
   Nevertheless, I need to sketch the various classes of mass
measurements to outline how they enter the evaluation of masses and how
they interfere with each other.

   Generally a mass measurement can be obtained by establishing
an energy relation between the mass we want to determine and a well
known nuclidic mass.
   This energy relation is then expressed in electron-volts (eV).
   Mass measurements can also be obtained as an inertial mass from its
movement characteristics in an electro-magnetic field.
   The mass, thus derived from a ratio of masses, is then expressed in
`unified atomic mass' (u) (cf.~Section~\ref{sect:masspec}),
or its sub-unit, $\mu$u.
   Two units are thus used in the present work.

   The mass unit is defined, since 1960, by 1\,u $= \Ma(^{12}$C$)/12$, one
twelfth of the mass of one free atom of Carbon-12 in its atomic and
nuclear ground-states.
   Before 1960, as Wapstra once told me, two mass units were defined:
   the physical one $^{16}$O$/16$,
   and the chemical one which considered one sixteenth of the average
mass of a standard mixture of the three stable isotopes of oxygen.
   Physicists could not convince the chemists to drop their unit;
   ``The change would mean millions of dollars in the sale of all
chemical substances", said the chemists, which is indeed true!
   Joseph~H.E.~Mattauch, the American chemist Truman~P.~Kohman and
Aaldert~H.~Wapstra \cite{58kohm} then calculated that, if
$^{12}$C$/12$ was chosen, the change would be ten times smaller for
chemists, and in the opposite direction \ldots
   That lead to unification;
   `u' stands therefore, officially, for `unified mass unit'!
   Let us mention to be complete that the chemical mass spectrometry
community (e.g. bio-chemistry, polymer chemistry) widely use the dalton
(symbol Da, named after John Dalton \cite{dalton}), which allows to
express the number of nucleons in a molecule.
   It is thus not strictly the same as `u'.

   The energy unit is the electronvolt.
   Until recently, the relative precision of $M-A$ expressed in keV was,
for several nuclides, less good than the same quantity expressed in mass
units.
   The choice of the volt for the energy unit (the electronvolt) is not
evident.
   One might expect use of the {\em international} volt V, but one can
also choose the {\em standard} volt V$_{90}$ as maintained in national
laboratories for standards and defined by adopting an exact value for
the constant ($2e/h$) in the relation between frequency and voltage in
the Josephson effect.
   In the 1999 table of standards \cite{99mohr}:
$2e/h=483597.9$\,(exact)\,GHz/V$_{90}$.
   An analysis by Cohen and Wapstra \cite{83coh4} showed that all
precision measurements of reaction and decay energies were calibrated in
such a way that they can be more accurately expressed in V$_{90}$.
   Also, the precision of the conversion factor between mass units and
{\em standard} volts V$_{90}$ is more accurate than that between it
and {\em international} volts~V:
   \begin{eqnarray*}
   1 \, {\rm u} & = & 931\,494.009\,0   \pm  0.007\,1  \0 {\rm keV}_{90}  \\
   1 \, {\rm u} & = & 931\,494.013\,\0  \pm  0.037\,\0 \0 {\rm keV}
   \end{eqnarray*}
   The reader will find more information on the energy unit, and also
some historical facts about the electronvolt, in the {\sc Ame2003}
\cite{Ame03}, page~134.

   \subsection{The experimental data} \label{sect:src}

   In this section we shall examine the various types of experimental
information on masses and see how they enter the {\sc Ame}.

   \subsubsection{Reaction energies}

   The energy absorbed in a nuclear reaction is directly derived from
the Einstein's relation $E=mc^2$.
   In a reaction A(a,b)B requiring an energy $Q_r$ to occur, the energy
balance writes:
   \begin{equation}\label{equ:qr}
   Q_r = \Ma_{\rm A} + \Ma_{\rm a} - \Ma_{\rm b} - \Ma_{\rm B}
   \end{equation}
   This reaction is often endothermic, that is $Q_r$ is negative,
requiring input of energy to occur.
   Other nuclear reactions may release energy.
   This is the case, for example, for thermal neutron-capture reactions
(n,$\gamma$) where the (quasi)-null energetic neutron is absorbed and
populates levels in the continuum of nuclide `B' at an excitation energy
exactly equal to $Q_r$.
   With the exception of some reactions between very light nuclides, the
the masses of the projectile `a' and of the ejectile `b' are
known with a much higher accuracy than those of the target `A', and of
course the residual nuclide `B'.
   Therefore Eq.\,\ref{equ:qr} reduces to a linear combination of the
masses of two nuclides:
   \begin{equation}\label{equ:qdq}
   \Ma_{\rm A} - \Ma_{\rm B} = q \pm dq
   \end{equation}
   where $q=Q_r -\Ma_{\rm a} + \Ma_{\rm b}$.

   A nuclear reaction usually deals with stable or very-long-lived target
`A' and projectile `a', allowing only to determine the mass of a residual
nuclide `B' close to stability.
   Nowadays with the availability of radioactive beams, interest in
reaction energy experiments is being revived.

   It is worth mentioning in this category the very high accuracies
attainable with (n,$\gamma$) and (p,$\gamma$) reactions.
   They play a key-r\^ole in providing many of the most accurate mass
differences, and help thus building and strengthening the
``backbone"\footnote{
      the `backbone' is the ensemble of nuclides along the line of
      stability in a diagram of atomic number $Z$ versus neutron number
      $Z$ \cite{Ame77d}.
      The energy relations among nuclides in the backbone are multiple.
   } of masses along the valley of $\beta$-stability, and determine
neutron separation energies with high precision\footnote{
      The number of couples of nuclides connected by
      (n,$\gamma$) reactions with an accuracy of 0.5\,keV or better was
      243 in {\sc Ame2003}, against 199 in {\sc Ame93}, 128 in {\sc Ame83}
      and 60 in the 1977 one.
      The number of cases known to better than 0.1\,keV was 100 in
      {\sc Ame2003}, 66 in {\sc Ame93} and 33 in {\sc Ame83}.
      Several reaction energies of (p,$\gamma$) reactions are presently
      ({\sc Ame93})known about as precisely (25 and 8 cases with
      accuracies better than 0.5\,keV and 0.1\,keV respectively).
   }.

   Also very accurate are the self-calibrated reaction energy
measurements using spectrometers.
   When measuring the difference in energy between the spectral lines
corresponding to reactions A(a,b)B and C(a,b)D with the same
spectrometer settings \cite{87Koslo} one can reach accuracies better
than 100~eV.
   Here the measurement can be represented by a linear combination of
the masses of four nuclides:
   \begin{equation}\label{equ:dqr}
   \delta Q_r = \Ma_{\rm A} - \Ma_{\rm B} - \Ma_{\rm C} + \Ma_{\rm D}
   \end{equation}

   The most precise reaction energy is the one that determined the mass
of the neutron from the neutron-capture energy of $^{1}$H at the
{\sc Ill} \cite{99Kessler}.
   The $^{1}$H(n,$\gamma$)$^{2}$H established a relation between the
masses of the neutron, of $^{1}$H and of the deuteron with the
incredible precision of 0.4~eV.

   \subsubsection{Disintegration energies}

   Disintegration can be considered as a particular case of reaction,
where there is no incident particle.
   Of course, here the energies $Q_\beta$, $Q_\alpha$ or $Q_p$ are
almost always positive, i.e. these particular reactions are exothermic.
   For the A($\beta^-$)B, A($\alpha$)B or A(p)B disintegrations\footnote{
     The drawing for $\alpha$-decay is taken from the educational Web
     site of the Lawrence Berkeley Laboratory: http://www.lbl.gov/abc/.
   }, one can write respectively:
   \begin{eqnarray}
     Q_{\beta^-}    & = & \Ma_{\rm A} - \Ma_{\rm B}  \\
     Q_{\alpha}\,\, & = & \Ma_{\rm A} - \Ma_{\rm B} - \Ma_\alpha  \\
     Q_{p}     \,\, & = & \Ma_{\rm A} - \Ma_{\rm B} - \Ma_{\rm p}
   \end{eqnarray}
   \begin{figure*}[htb]   
   \begin{center}
   \includegraphics[width=11 cm]{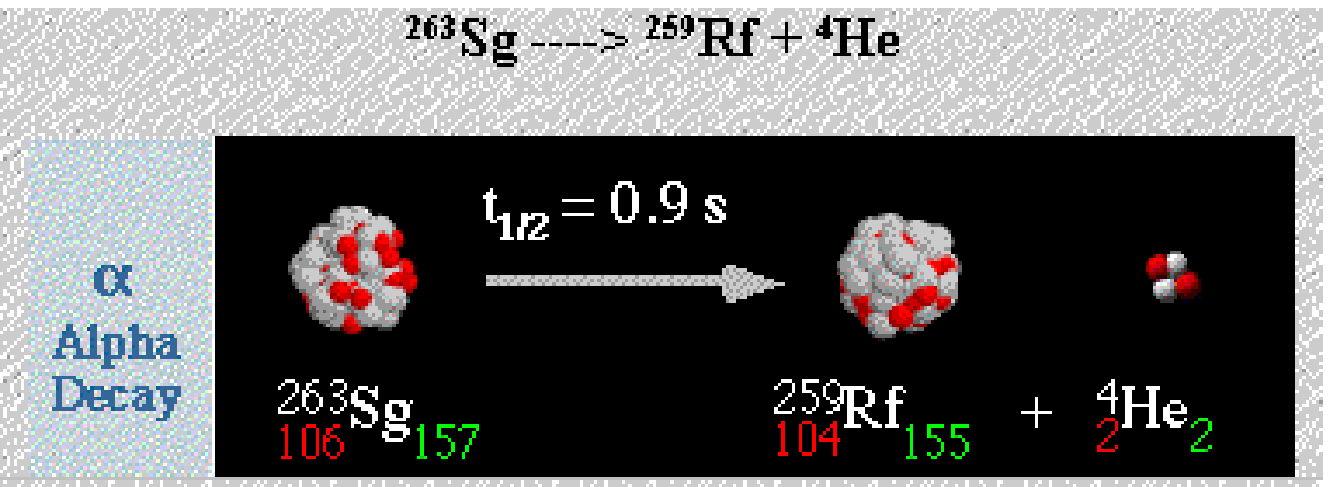}
   \label{fig:alpha}
   \end{center}
   \end{figure*}

   These measurements are very important because they allow deriving
masses of unstable or very unstable nuclides.

   This is more specially the case for the proton decay of nuclides at
the drip-line, in the medium-$A$ region \cite{96Davids}.
   They allow a very useful extension of the systematics of proton
binding energies.
   But in addition they give in several cases information on excitation
energies of isomers.
   This development is one more reason why we have to give more
attention to relative positions of isomers than was necessary in
earlier evaluations.

   $\alpha$-decays have permitted to determine the masses of the heavy
nuclides.
   Moreover, the time coincidence of $\alpha$ lines in a decaying chain
allows very clear identification of the heaviest ones.
   Quite often, and more particularly for even-even nuclides, the
measured $\alpha$-decay energies yield quite precise mass differences
between parent and daughter nuclide.

   \subsubsection{Mass Spectrometry} \label{sect:masspec}

   Mass-spectrometric determination of atomic masses are often called
`direct' mass measurements because they are supposed to determine not an
energy relation between two nuclides, but directly the mass of the
desired one.
   In principle this is true, but only to the level of accuracy of the
parameter of the spectrometer that is the least well known, which is
usually the magnetic field in which the ions move.
   It follows that the accuracy in such absolute direct mass
determination is very poor.

   This is why, in all precise mass measurements, the mass of an unknown
nuclide is always compared, in the same magnetic field, to that of a
reference nuclide.
   Thus, one determines a ratio of masses, where the value of the
magnetic field cancels, leading to a much more precise mass
determination.
   As far as the {\sc Ame} is concerned, here again we have a mass
relation between two nuclides.

   One can distinguish three sub-classes in the class of mass
measurement by mass-spectrometry (see also \cite{Orr-Lepine}):
   \begin{enumerate}
   \item
   Classical mass-spectrometry, where the electromagnetic deflection
plays the key-r\^ole.
   More exactly, the two beams corresponding to the ion of the
investigated nuclide and to that of the reference are forced to follow
the same path in the magnetic field.
   The ratio of the voltages of some electrostatic devices that make
this condition true determines the ratio of masses.
   These voltages are determined either from the values of resistors in
a bridge \cite{63Ries} or directly from a precision voltmeter
\cite{63Barber}.

   \item
   Time-of-Flight spectrometry, where one measures simultaneously the
momentum of an ion (from its magnetic rigidity $B\rho$) and its velocity
(from the time of flight along a path of well-determined length)
\cite{Gillibert-Vieira}.
   Calibration in this type of experiment requires a large set of
reference masses, so that the {\sc Ame} cannot establish a simple
relation between two nuclides.
   Nevertheless, the calibration function thus determined, together with
its contribution to the error is generally well accounted for.
   The chance is small that recalibration might be necessary.
   In case it appears to be so in some future, one could consider a
global re-centering of the published values.
   It is interesting to note that Time-of-Flight spectrometers can be
also set-up in cyclotrons \cite{96Chartier} or in storage rings
\cite{92Trotscher}.

   \item
   Cyclotron Frequency, when measured in a homogeneous magnetic field,
yields mass value of very high precision due to the fact that frequency
is the physical quantity that can be measured with the highest accuracy
with the present technology.
   Three types of spectrometers follow this principle:
   \begin{itemize}
   \item
   the {\em Radio-Frequency Mass Spectrometer} (Fig.\,\ref{fig:srf})
invented by L.G.~Smith \cite{56Smith+58} where the measurement is
obtained in-flight, as a transmission signal, in only one turn;
   \begin{figure}[htb]   
   \begin{center}
   \includegraphics[height=4.3 cm]{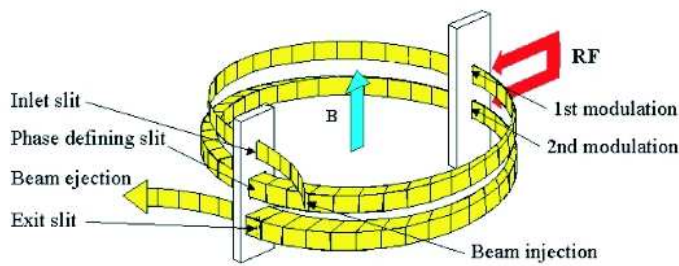}
   \caption[]{\footnotesize
   Principle of the {\em Radio-Frequency Mass Spectrometer}.
   Ions make two turns following a helicoidal path in a homogeneous
field $\vec B$.
   Two RF excitations are applied at one turn interval.
   Only ions for which the two excitations are in opposite phase (and
then cancel) will exit the spectrometer and be detected.
   Typical diameter of the helix is 0.5-1 meter.
   This scheme is from the {\sc Mistral} Web site: http://csnwww.in2p3.fr/groupes/massatom/.
   }\label{fig:srf}
   \end{center}
   \end{figure}

   \item
   the {\em Penning Trap Spectrometer} (Fig.\,\ref{fig:trap}) where the
ions are stored for 0.1--2 seconds to interact with a
radio-frequency excitation signal \cite{81Schwinberg}; and
   \begin{figure}[htb]   
   \begin{center}
   \includegraphics[width=5 cm]{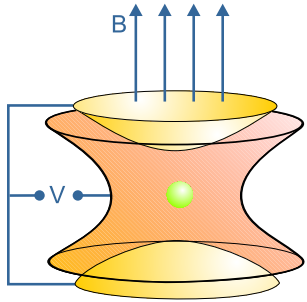}                         
   \caption[]{\footnotesize
   Principle of the {\em Penning Trap Spectrometer}.
   Ions follow a cyclotron motion in the horizontal plane due to $\vec
B$ and cannot escape axially due to repulsion by the end-cup electrodes.
   The ring electrode is split to allow RF excitation.
   Typical inner size is 1-2 cm.
   This scheme is from the {\sc Isoltrap} Web site: http://cern.ch/isoltrap/.
   }\label{fig:trap}
   \end{center}
   \end{figure}

   \item
   the {\em Storage Ring Spectrometer} where the ions are stored and the
ion beam cooled, while a metallic probe near the beam picks up the
generated Schottky noise (a signal induced by a moving charge)
\cite{94Franzke}.
   \end{itemize}
   \end{enumerate}

   Penning traps, as well as storage rings and the {\sc Mistral} on-line
Smith-type spectrometer, are now also used for making mass measurements
of many nuclides further away from the line of stability.
   As a result, the number of nuclides for which experimental mass
values are now known is substantially larger than in the previous
atomic mass tables.
   These mass-spectrometric measurements of exotic nuclides are often
made with resolutions that do not allow separation of isomers.
   Special care is needed to derive the mass value for the
ground-state (cf.~Section~\ref{sect:Nub}).

   Nowadays, several mass measurements are made on fully (bare nuclei)
or almost fully ionized atoms.
   Then, a correction must be made for the total binding energy of all
removed electrons $B_e(Z)$.
   They can be found in the table for calculated total atomic binding
energy of all electrons of Huang et al. \cite{76hua}.
   Unfortunately, the precision of the calculated values $B_e (Z)$ is
not clear; this quantity (up to 760~keV for $_{92}$U)
cannot be measured easily.
   Very probably, its precision for $_{92}$U is rather better than the
2~keV accuracy with which the mass of, e.g., $^{238}$U is known.
   A simple formula, approximating the results of \cite{76hua}, is given
in the review of Lunney, Pearson and Thibault~\cite{03rmp}:
  \begin{equation}
      B_{el}(Z) = 14.4381\, Z^{2.39} + 1.55468 \times 10^{-6}\, Z^{5.35}
      \:\mbox{eV}
  \end{equation}

\begin{figure}[htb]   
   \begin{center}
   \includegraphics[height=15 cm,bb= 80 32 574 742,clip,angle=270]{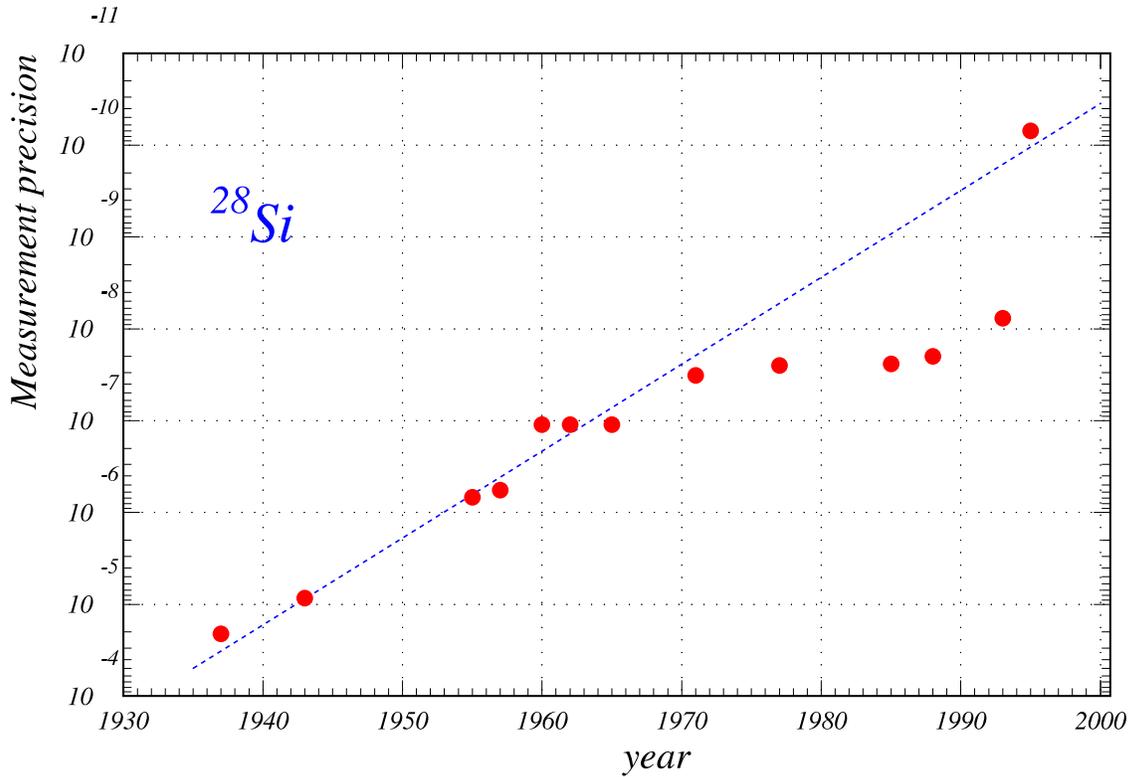}
   \caption[]{\footnotesize
   Our knowledge of the mass of $^{28}$Si has increased by one order of
magnitude per decade since 1935.
   Extrapolating this tendency, one could expect that we will reach a
$10^{-12}$ accuracy in 2015.
   }\label{fig:28si}
   \end{center}
   \end{figure}

   Penning traps have allowed to reach incredible accuracies in the
measurement of masses.
   If one observes the increase of accuracies over the last seven or
eight decades, for example on $^{28}$Si, see Fig.\,\ref{fig:28si}, one
would see that after a regular increase of one order of magnitude every
ten years until 1970, the mass accuracy of $^{28}$Si seemed to have
reached a limit at the level of $5\times10^{-7}$.
   Until the arrival of high precision Penning traps, that allowed to
catch up with the previous tendency and yielded an accuracy slightly
better than $10^{-10}$ in 1995 \cite{95mit}.
   Such precision measurements with Penning traps have considerably
improved the precision in our knowledge of atomic mass values along the
backbone.

   \subsection{Data evaluation in the {\sc Ame}} \label{sect:eval}

   The evaluation of masses share with most other evaluations many
procedures.
   However, the very special character in the treatment of data in the
mass evaluation is, as said above (header of the present Section), that
all measurements are relative measurements.
   Each experimental datum will be thus represented by a link connecting
two or three nuclei (cf.~Section~\ref{sect:correl}).
   The set of connections results in a complex canvas where data of
different type and origin are entangled.
   Here lies the very challenge to extract values of masses from the
experiments.
   The counterpart is that the overdetermined data system will allow
cross-checks and studies of the consistencies within this system.
   The other help to the evaluator will be the property of regularity of
the surface of masses that will be described in the last section of this
lecture.

   The first step in the evaluation of data is to make a compilation,
i.e. a collection of all the available data.
   This collection must include the `hidden' data: a paper does not
always say clearly that some of the
information it contains is of interest for mass measurement.
   The collection includes also even poorly documented datum, which
is labelled accordingly in the {\sc Ame} files.

   The second step is the critical reading, which might include:
   \begin{enumerate}
   \item the evaluation or re-evaluation of the calibration procedures,
the calibrators, and of the precisions of the measurements;
   \item spectra examination: peaks position and relative intensities,
peaks symmetry, quality of the fit;
   \item search for the {\sc primary} information, in the data, which do
not necessarily appear always as clearly as they should.
   (i.e. in cases the authors combined the original result with other
data, to derive a mass value, the {\sc Ame} should retain only the
former).
   It also happened that the authors gave only the derived mass value.
   Then the evaluator has to reconstruct the original result and ask for
authors' confirmation.
   \end{enumerate}

   The third step in the data evaluation will be to compare the results
of the examined work to earlier results if they exist (either directly,
or through a combination of other data).
   If there are no previous results, comparison could be done with
estimates from extrapolations, exploiting the above mentioned regularity
of the mass surface (cf.~Section\,~\ref{sect:msreg}), or to estimates
from mass models or mass formulae.

   Finally, the evaluator often and on has to establish a dialog with
the authors of the work, asking for complementary information when
necessary, or suggesting different analysis, or suggesting new
measurements.

   The new data can now enter the data-file as one line.
   For example, for the electron capture of $^{205}$Pb, the evaluator
enters:
   {\small\begin{verbatim}
205 890816000c1 B  78Pe08   41.4   1.1   205Pb(e)205Tl    0.525  0.008 LM
   \end{verbatim}}
   \noindent
   where besides a 14 digits ID-number, there is a flag (as described in
Ref.\,\cite{Ame03}, p.~184), here `B', then the {\sc Nsr} reference-code
\cite{nsr} for the paper `78Pe08' where the data appeared, the value for
the $Q$ of the reaction with its error bar (41.4 $\pm$ 1.1 keV), and the
reaction equation, where `e' stands for electron-capture.
   The information in the last columns says that this datum has been
derived from the intensity ratio (0.525 $\pm$ 0.008) of the L and M
lines in electron capture.
   The evaluator can add as many comment lines as necessary, following
this data line, for other information he judges useful for exchange with
his fellow evaluator.
   Some of these comments, useful for the user of the mass tables, will
appear in the {\sc Ame} publication.

   \subsection{Data treatment} \label{sect:treat}

   In this section, we shall first see how the network of data is built,
then how the system of data can be reduced.
   In the third and fourth subsections, I shall describe shortly the
least-squares method used in the {\sc Ame} and the computer program that
will decode data and calculate the adjusted masses.
   A fifth part will develop the very important concept of
`Flow-of-Information' matrix.
   Finally, I shall explain how checking the consistency of data (or of
sub-group of data) can help the evaluator in his judgment.

   \subsubsection{Data entanglement - Mass Correlations} \label{sect:correl}

\begin{figure}[htb]   
   \begin{center}
   \rotatebox{-90}{\includegraphics[bb=32 136 562 720,clip,height=15cm]{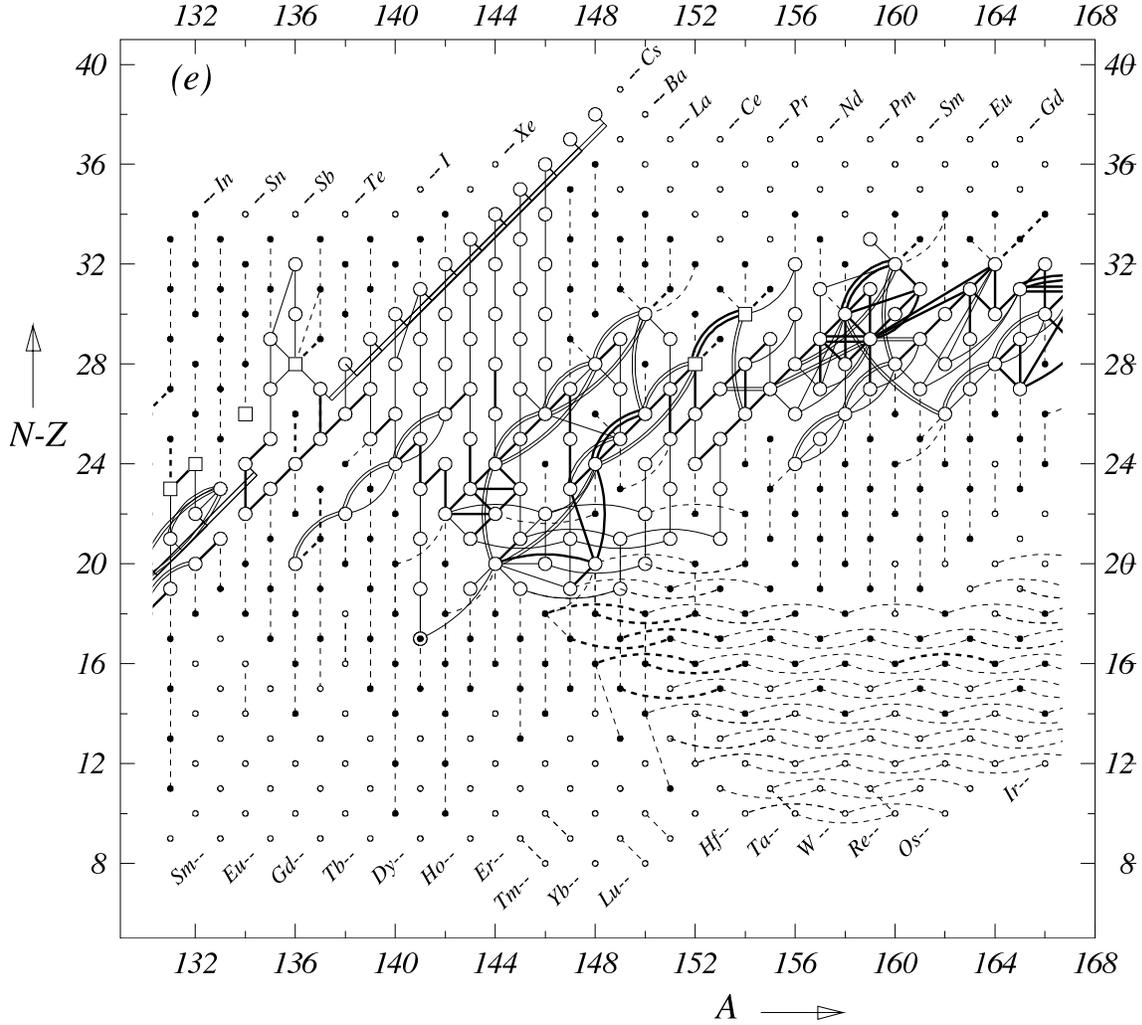}}
   \caption[]{\footnotesize
   Diagram of connections for the experimental data.
   Each symbol represents one nuclide and each line represents one piece
of data connecting two nuclides.
   When a nuclide is connected to Carbon-12 (often the case for mass
spectrometry), it is represented by a square symbol.
   }\label{fig:conns}
   \end{center}
   \end{figure}

   We have seen in Section~\ref{sect:src} that all mass measurements are
{\sc relative} measurements.
   Each experimental piece of data can be represented by a link between
two, sometimes three, and more rarely four nuclides.
   As mentioned earlier, assembling these links produces an extremely
entangled network.
   A part of  this network can be seen in Fig.\,\ref{fig:conns}.
   One notices immediately that there are two types of symbols, the
small and the large ones.
   The small ones represent the so-called {\sc secondary} nuclides;
while the nuclides with large symbols are called {\sc primary}.
   Secondary nuclides are represented by full small circles if their
mass is determined experimentally, and by empty ones if estimated from
trends in systematics.
   Secondary nuclides are connected by {\sc secondary} data, represented
by dashed lines.
   A chain of dashed lines is at one end free, and at the other end
connected to one unique\footnote{
      Sometimes a chain of secondary nuclides can be free at both ends.
      These nuclides have no connection to the backbone of known masses,
      but are connected to each other by $\alpha$-chains of sometime high or
      very high precision.
      The chain is floating and no experimental mass can be derived.
      The evaluator makes an estimate for one of the masses in the chain in
      order to have it fixed.
      These non-experimental masses are all quoted as (`systematics') in
      the Tables.
   } primary nuclide (large symbol).
   This representation means that all secondary nuclides are determined
uniquely by the chain of secondary connections going down to a primary
nuclide.
   The latter are multiply determined and enter thus the entangled
canvas.
   They are inter-connected by {\sc primary} data, represented by full
lines.

   We see immediately from Fig.\,\ref{fig:conns} that the mass of a
primary nuclide cannot be determined straightforwardly.
   One may think of making an average of the values obtained from all
links, but such a recipe is erroneous because the other nuclides on
which these links are built are themselves inter-connected, thus not
independent.
   In other words these {\sc primary} data, connecting
the primary nuclides, are correlated, and the correlation coefficients
are to be taken into account.

   Caveat: the word {\em primary} used for these nuclides and for the
data connecting them does not mean that they are more important than the
others, but only that they are subject to the special treatment below.
   The labels {\em primary} and {\em secondary} are not intrinsic
properties of data or masses.
   They may change from primary to secondary or reversely when other
information becomes available.

   \subsubsection{Compacting the set of data} \label{sect:cmpct}

   We have seen that {\em primary} data are correlated.
   We take into account these correlations very easily with the help of
the least-squares method that will be described below.
   The {\em primary} data will be improved in the adjustment, since each
will benefit from all the available information.

   {\em Secondary} data will remain unchanged; they do not contribute to
$\chi^2$.
   The masses of the secondary nuclides will be derived directly by
combining the relevant adjusted primary mass with the secondary datum or
data.
   This also means that secondary data can easily be replaced by new
information becoming available (but one has to watch since the
replacement can change other secondary masses down the chain as seen
from the diagram Fig.\,\ref{fig:conns}).

   We define {\sc degrees} for {\em secondary} masses and
{\em secondary} data.
   They reflect their distances along the chains connecting them to the
network of primaries; they range from 2 to 16.
   Thus, the first secondary mass connected to a primary one will be a
mass of degree 2, and the connecting datum will be a datum of degree 2
too.
   Degree 1 is for primary masses and data.

   Before treating the primary data by the least-squares method, we try
as much as possible to reduce the system, but without allowing any loss
of information.
   One way to do so is to {\sc pre-average} identical data: two or more
measurements of the same physical quantities can be replaced by their
average value and error.
   Also the so-called {\sc parallel} data can be pre-averaged: they are
data that give essentially values for the mass difference between the
same two nuclides, e.g. $^9$Be($\gamma$,n)$^8$Be, $^9$Be(p,d)$^8$Be,
$^9$Be(d,t)$^8$Be and $^9$Be($^3$He,$\alpha$)$^8$Be.
   Such data are represented together, in the main least-squares
calculation, by one of them carrying their average value.
   If the $Q$ data to be pre-averaged are strongly conflicting, i.e. if
the consistency factor (or Birge ratio, or normalized $\chi$)
   \begin{equation}\label{equ:kin}
   \chi_n = \sqrt{\frac{\chi^2}{Q-1}}
   \end{equation}
   resulting in the calculation of the pre-average is greater than 2.5,
the (internal) error $\sigma_i$ in the average is multiplied by the
Birge ratio ($\sigma_e=\sigma_i\times \chi_n$).
   The quantity $\sigma_e$ is often called the `external' error.
   However, this treatment is not used in the very rare cases where the
errors in the values to be averaged differ too much from one another,
since the assigned errors loose any significance (three cases in
{\sc Ame'93}).
   We there adopt an arithmetic average and the dispersion of values as
error, which is equivalent to assigning to each of these conflicting data
the same error.

   In {\sc Ame'93}, 28\% of the 929 cases in the pre-average had values
of $\chi_n$ beyond unity, 4.5\% beyond two, 0.7\% beyond 3 and only one
case beyond 4, giving a very satisfactory distribution overall.
   With the choice above of a threshold of $\chi_n^0$=2.5 for the Birge
ratio, only 1.5\% of the cases are concerned by the multiplication by
$\chi_n$.
   As a matter of fact, in a complex system like the one here, many
values of $\chi_n$ beyond 1 or 2 are expected to exist, and if errors
were multiplied by $\chi_n$ in all these cases, the $\chi^2$-test on the
total adjustment would have been invalidated.
   This explains the choice made in the {\sc Ame} of a rather high
threshold ($\chi_n^0=2.5$), compared e.g. to $\chi_n^0$=2 recommended by
Woods and Munster \cite{88Woods} or, even, $\chi_n^0$=1 used in a
different context by the Particle Data Group \cite{98pdg}, for departing
from the rule of internal error of the weighted average (see also
\cite{92Wapstra}).

   Another method to increase the meaning of the final $\chi^2$ is to
exclude data with weights at least a factor 10 less than other data, or
combinations of other data giving the same result.
   They are still kept in the list of input data but labelled
accordingly; comparison with the output values allows to check that this
procedure did not have unwanted consequences.

   The system of data is also greatly reduced by replacing data with
isomers by an equivalent datum for the ground-state, if a $\gamma$-ray
energy measurement is available from the {\sc Nndc}
(cf.~Section~\ref{sect:Nub}).
   Excitation energies from such $\gamma$-ray measurements are normally
far more precise than reaction energy measurements.

   Typically, we start from a set of 6000 to 7000 experimental data
connecting some 3000 nuclides.
   After pre-averaging, taking out the data with very poor accuracy and
separating the secondary data, we are left with a system of 1500 primary
data for 800 nuclides.

   \subsubsection{Least-squares method}  \label{sect:lsq}

   Each piece of data has a value $q_i \pm dq_i$ with the accuracy
$dq_i$ (one standard deviation) and makes a relation between 2, 3 or 4
masses with unknown values $m_{\mu}$.
   An overdetermined system of $Q$ data to $M$ masses ($Q > M$) can be
represented by a system of $Q$ linear equations with $M$ parameters:
   \begin{equation}\label{equ:linear}
   \sum_{\mu=1}^M k_i^{\mu} m_{\mu} = q_i \pm dq_i
   \end{equation}
   (a generalization of Eq.\,\ref{equ:qdq} and Eq.\,\ref{equ:dqr}),
   e.g. for a nuclear reaction $A(a,b)B$ requiring an energy $q_i$ to
occur, the energy balance writes:
   \begin{equation}\label{equ:qi}
   m_{\rm A} + m_{\rm a} - m_{\rm b} - m_{\rm B} = q_i \pm dq_i
   \end{equation}
   thus, $k_i^{\rm A}=+1$, ~ $k_i^{\rm a}=+1$, ~ $k_i^{\rm B}=-1$ ~ and ~ $k_i^{\rm b}=-1$.

   In matrix notation, ${\bf K}$ being the $(Q,M)$ matrix of
coefficients, Eq.\,\ref{equ:linear} writes:
   $ {\bf K} \vert m\rangle = \vert q\rangle $.
   Elements of matrix ${\bf K}$ are almost all null:
   e.g. for reaction $A(a,b)B$, Eq.\,\ref{equ:qdq} yields a line of ${\bf K}$ with
only two non-zero elements.

   We define the diagonal weight matrix ${\bf W}$ by its elements
$w_i^i= 1 / (dq_idq_i)$.
   The solution of the least-squares method leads to a very simple
construction:
   \begin{equation}
   {\bf ^tKWK} \vert m\rangle = {\bf ^tKW} \vert q\rangle
   \end{equation}
   the {\sc normal} matrix ${\bf A}={\bf ^tKWK}$ is a square matrix of
order $M$, positive-definite, symmetric and regular and hence invertible
\cite{61Linnik}. Thus the vector $\vert\overline m\rangle$ for the adjusted
masses is:
   \begin{equation}
   \vert \overline m\rangle = {\bf {A^{-1}} ~ ^tKW} \vert q\rangle \nr {\rm or} \nr
   \vert \overline m\rangle = {\bf R} \vert q\rangle
   \end{equation}
   The rectangular $(M,Q)$ matrix ${\bf R}$ is called the {\sc response}
matrix.

   The diagonal elements of ${\bf A^{-1}}$ are the squared errors on the
adjusted masses, and the non-diagonal ones $(a^{-1})_{\mu}^{\nu}$
are the coefficients for the correlations between masses $m_{\mu}$
and $m_{\nu}$.

   \subsubsection{The {\sc Ame} computer program}

   The four phases of the {\sc Ame} computer program perform the
following tasks:
   \begin{enumerate}
   \item
   decode and check the data file;
   \item
   build up a representation of the connections between masses, allowing
thus to separate primary masses and data from secondary ones and then to
reduce drastically the size of the system of equations to be solved,
without any loss of information;
   \item
   perform the least-squares matrix calculations (see above); and
   \item
   deduce the atomic masses,
   the nuclear reaction and separation energies,
   the adjusted values for the input data,
   the {\em influences} of data on the primary masses described in next
section,
   and display information on the inversion errors,
   the correlations coefficients,
   the values of the $\chi^2$ (cf.~Section~\ref{sect:chi2}), and
   the distribution of the normalized deviations $v_i$.
   \end{enumerate}

   \subsubsection{Flow-of-Information}

   One of the most powerful tools in the least-squares calculation
described above is the flow-of-information matrix.
   This matrix allows to trace back, in the least-squares method, the
contribution of each individual
piece of data to each of the parameters (here the atomic masses).
   The {\sc Ame} uses this method since 1993.

   The flow-of-information matrix ${\bf F}$ is defined as follows:
   ${\bf K}$, the matrix of coefficients, is a rectangular $(Q,M)$
matrix,
   the transpose of the response matrix ${\bf ^tR}$ is also a $(Q,M)$
rectangular one.
   The $(i,\mu)$ element of ${\bf F}$ is defined as the product of
the corresponding elements of ${\bf ^tR}$ and of ${\bf K}$.
   In reference \cite{86nim} it is demonstrated that such an element
represents the {\em ``influence''} of datum $i$ on parameter (mass)
$m_{\mu}$.
   A column of ${\bf F}$ thus represents all the contributions brought
by all data to a given mass $m_{\mu}$, and a line of ${\bf F}$
represents all the influences given by a single piece of data.
   The sum of influences along a line is the {\em ``significance''} of
that datum.
   It has also been proven \cite{86nim} that the influences and
significances have all the expected properties, namely that the sum of
all the influences on a given mass (along a column) is unity, that the
significance of a datum is always less than unity and that it always
decreases when new data are added.
   The significance defined in this way is exactly the quantity obtained
by squaring the ratio of the uncertainty on the adjusted value over that
on the input one, which is the recipe that was used before the discovery
of the ${\bf F}$ matrix to calculate the relative importance of data.

   A simple interpretation of influences and significances can be
obtained in calculating, from the adjusted masses and
Eq.\,\ref{equ:linear}, the adjusted data:
   \begin{equation}\label{equ:adj}
    \vert \overline q\rangle = {\bf KR} \vert q\rangle.
   \end{equation}
   The $i^{th}$ diagonal element of ${\bf KR}$ represents then the
contribution of datum $i$ to the determination of $\overline{q_i}$ (same
datum): this quantity is exactly what is called above the {\em
significance} of datum $i$.
   This $i^{th}$ diagonal element of ${\bf KR}$ is the sum of the
products of line $i$ of ${\bf K}$ and column $i$ of ${\bf R}$.
   The individual terms in this sum are precisely the {\em influences}
defined above.

   The flow-of-information matrix ${\bf F}$, provides thus insight on
how the information from datum $i$ flows into each of the masses
$m_{\mu}$.

   \subsubsection{Consistency of data} \label{sect:chi2}     

   The system of equations being largely over-determined ($Q>>M$) offers
the evaluator several interesting possibilities to examine and judge the
data.
   One might for example examine all data for which the adjusted values
deviate importantly from the input ones.
   This helps to locate erroneous pieces of information.
   One could also examine a group of data in one experiment and check if the
errors assigned to them in the experimental paper were not underestimated.

   If the precisions $dq_i$ assigned to the data $q_i$ were indeed all
accurate, the normalized deviations $v_i$ between adjusted
$\overline q_i$ and input $q_i$ data (cf.~Eq.\,\ref{equ:adj}),
    $v_i = (\overline q_i - q_i)/dq_i$,
   would be distributed as a gaussian function of standard deviation
$\sigma=1$, and would make $\chi^2$:
   \begin{equation}\label{equ:chi2}
   \chi^2 = \sum_{i=1}^Q \left(\frac{\overline q_i - q_i}{dq_i}\right)^2
   \nr {\rm or} \nr
   \chi^2 = \sum_{i=1}^Q v_i^2
   \end{equation}
   equal to $Q-M$, the number of degrees of freedom, with a precision of
$\sqrt{2(Q-M)}$.

   One can define as above the {\sc normalized chi}, $\chi_n$ (or
`consistency factor' or `Birge ratio'): $\chi_n=\sqrt{\chi^2/(Q-M)}$
for which the expected value is $1\pm1/\sqrt{2(Q-M)}$.

   For our current {\sc Ame2003} example of 1381 equations with 847
parameters, i.e. 534 degrees
   of freedom, the theoretical expectation value for $\chi^2$ should be
534$\pm$33 (and the theoretical $\chi_n=1\pm0.031$).
   The total $\chi^2$ of the adjustment is actually 814; this means
that, in the average, the errors in the input values have been
underestimated by 23\%, a still acceptable result.
   In other words, the experimentalists measuring masses were, on
average, too optimistic by 23\%.
   The distribution of the $v_i$'s (the individual contributions to
$\chi^2$, as defined in Eq.\,\ref{equ:chi2} is
also acceptable, with, in {\sc Ame2003}, 15\% of the cases beyond unity, 3.2\% beyond two,
and 8 items (0.007\%) beyond 3.

   The $\chi_n$ value was 1.062 in {\sc Ame'83} for $Q-M$=760 degrees
of freedom, 1.176 in {\sc Ame'93} for $Q-M=635$, and 1.169 in the
{\sc Ame'95} update for $Q-M$=622.

   Another quantity of interest for the evaluator is the {\sc partial
consistency factor}, $\chi_n^p$, defined for a (homogeneous) group of
$p$ data as:
   \begin{equation}\label{equ:pcf}
   \chi_n^p = \sqrt{\frac{Q}{Q-M} \0 \frac{1}{p} \0 \sum_{i=1}^p v_i^2}.
   \end{equation}
   Of course the definition is such that $\chi_n^p$ reduces to $\chi_n$
if the sum is taken over all the input data.

   One can consider for example the two main classes of data: the
reaction and decay energy measurements and the mass spectrometric data
(see Section\,\ref{sect:src}).
   The partial
consistency factors $\chi_n^p$ are respectively 1.269 and 1.160 for
energy measurements and for mass spectrometry data, showing that both
types of input data are responsible for the underestimated error of 23\%
mentioned above, with a better result for mass spectrometry data.

   One can also try to estimate the
average accuracy for 181 groups of data related to a given laboratory
and with a given method of measurement, by calculating their partial
consistency factors $\chi_n^p$.
   A high value of $\chi_n^p$ might be a warning on the validity of
the considered group of data within the reported errors.
   In general, in the {\sc Ame} such a situation is extremely rare,
because deviating data are cured before entering the `machinery' of the
adjustment, at the stage of the evaluation itself (see
Section~\ref{sect:eval}).
   On the average the experimental errors appear to be slightly
underestimated, with as much as 57\% (instead of expected 33\%) of the
groups of data having $\chi_n^p$ larger than unity.
   Agreeing better with statistics, 5.5\% of these groups are beyond
$\chi_n^p=2$.

   \subsection{Data requiring special treatment}

   It often happens that data require some special treatment before
entering the data-file (cf.~Section~\ref{sect:eval}).
   Such is the case of data given with asymmetric uncertainties, or when
information is obtained only as one lower and one upper limit, defining
thus a range of values.
   We shall examine these two cases.

   All errors entering the data-file must be one standard deviation
(1~$\sigma$) errors.
   When it is not the case, they must be converted to 1~$\sigma$ errors
to allow combination with other data.

   \subsubsection{Asymmetric errors}

\begin{figure}[htb]   
   \begin{center}
   \includegraphics[bb=100 50 510 250,clip,height=4.5 cm]{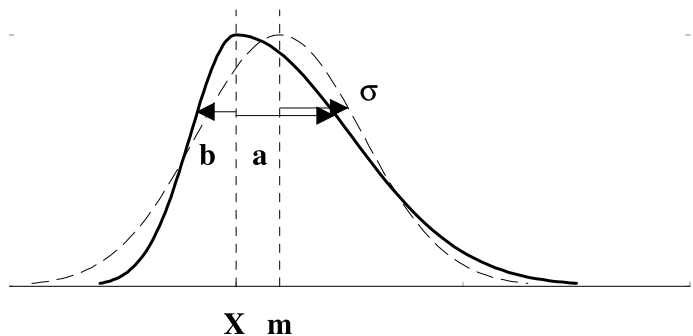}   
   \caption[]{\footnotesize
   An experimental result is represented by an asymmetric probability
density function (heavy solid line) with central value $X$ and errors
$+a$ and $-b$.
   This function is symmetrized as shown by the dashed line with its
center $m$ displaced by $0.64\cdot(a-b)$
   }\label{fig:asym}
   \end{center}
   \end{figure}

   Sometimes the precision on a measurement is not given as a single
number, like $\sigma$ (or $dq$ in Section~\ref{sect:lsq} above), but
asymmetrically $X^{+a}_{-b}$, as shown in Fig.\,\ref{fig:asym}.

   Such errors are symmetrized, before entering the treatment procedure.
   A rough estimate can be used: take the central value to be the
mid-value between the upper and lower 1$\sigma$-equivalent limits
$X+(a-b)/2$, and define the uncertainty to be the average of the two
uncertainties $(a+b)/2$.
   A better approximation is obtained with the recipe described in
Ref.\,\cite{Nubase03}.
   The central value $X$ is shifted to:
   \begin{equation}\label{equ:asym}
   X+0.64\cdot(a-b)
   \end{equation}
   and the precision $\sigma$ is:
   \begin{equation}\label{equ:uasym}
   \sigma^2 = (1 - \frac{2}{\pi}) \, (a-b)^2 + ab .
   \end{equation}

   In the appendix of Ref.\,\cite{Nubase03} one can find the
demonstration and discussion of Eq.\,\ref{equ:asym} and
Eq.\,\ref{equ:uasym}.

   \subsubsection{Range of values}

\begin{figure}[htb]   
   \begin{center}
   \includegraphics[height=4.5 cm]{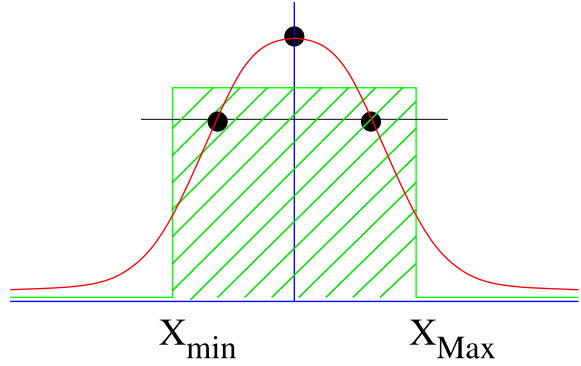}
   \caption[]{\footnotesize
   Experimental datum given as a range of values is represented by a
rectangular distribution of probabilities.
   }\label{fig:range}
   \end{center}
   \end{figure}

   Some measurements are reported as a range of values with most
probable lower and upper limits (Fig.\,\ref{fig:range}).
   They are treated as a uniform distribution of probabilities
\cite{82Audi}.
   The moments of this distribution yield a central value at the middle
of the range and a 1$\sigma$ uncertainty of 29\% of that range.

   \subsubsection{Mixture of spectral lines}

   ****  to be completed  ****

   \section{Regularity of the mass-surface} \label{sect:msreg}

\begin{figure}[htb]   
   \begin{center}
   \includegraphics[width=8 cm]{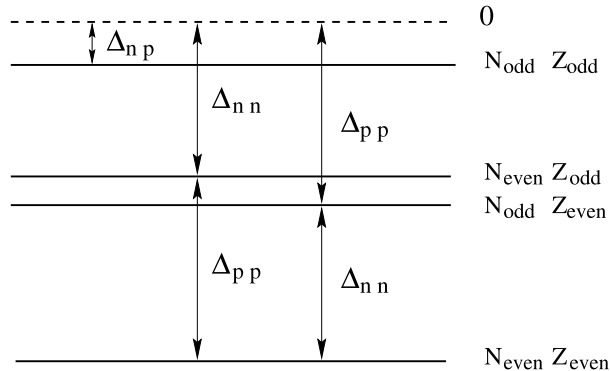}
   \caption[]{\footnotesize
   The surface of masses is split into four sheets.
   This scheme represents the pairing energies responsible for this
splitting.
   The zero energy surface is a purely hypothetical one for no pairing
at all among the last nucleons.
   }\label{fig:pairing}
   \end{center}
   \end{figure}

   When nuclear masses are displayed as a function of $N$ and $Z$,
one obtains a {\em surface} in a 3-dimensional space.
   However, due to the pairing energy, this surface splits into four
{\em sheets}.
   The even-even sheet lies lowest, the odd-odd highest, the other two
nearly halfway between as represented in Fig.\,\ref{fig:pairing}.
   The vertical distances from the even-even sheet to the odd-even and
even-odd ones are the proton and neutron pairing energies $\Delta_{pp}$
and $\Delta_{nn}$.
   They are nearly equal.
   The distances of the last two sheets to the odd-odd sheet are equal
to $\Delta_{nn}-\Delta_{np}$ and $\Delta_{pp}-\Delta_{np}$, where
$\Delta_{np}$ is the proton-neutron pairing energy due to the
interaction between the two odd nucleons, which are generally not in the
same shell.
    These energies are represented in Fig.\,\ref{fig:pairing},
where a hypothetical energy zero represents a nuclide with no pairing
among the last nucleons.

   Experimentally, it has been observed that:
   \begin{itemize}
   \item
   the four sheets run nearly parallel in all directions, which means
that the quantities $\Delta_{nn}$, $\Delta_{pp}$ and $\Delta_{np}$ vary
smoothly and slowly with $N$ and $Z$; and
   \item
   each of the mass sheets varies very smoothly also, but very
rapidly\footnote{
     smooth means continuous, non-staggering; smooth does not mean slow.
   }.
with $N$ and $Z$.
   The smoothness is also observed for first order derivatives (slopes,
cf.~Section~\ref{sect:deriv}) and all second order derivatives (curvatures
of the mass surface).
   They are only interrupted in places by sharp cusps or large
depressions associated with important changes in nuclear structure:
shell or sub-shell closures (sharp cusps), shape transitions
(spherical-deformed, prolate-oblate: depressions on the mass surface),
and the so-called `Wigner' cusp along the $N=Z$ line.
   \end{itemize}

   This observed regularity of the mass sheets in all places where no
change in the physics of the nucleus are known to exist, can be
considered as one of the {\sc basic properties} of the mass surface.
   Thus, dependable estimates of unknown, poorly known or questionable
masses can be obtained by extrapolation from well-known mass values on
the same sheet.
   Examination of previous such estimates, where the masses are now
known, shows that the method used, though basically simple, has a
good predictive power \cite{03rmp}.

   In the evaluation of masses the property of regularity and the
possibility to make estimates are used for several purposes:
   \begin{enumerate}
   \item
   {\em New Physics} - Any coherent deviation from regularity, in a
region $(N,Z)$ of some extent, could be considered as an indication that
some new physical property is being discovered.

   \item
   {\em Outliers} - However, if one single mass violates the systematic
trends, then one may seriously question the correctness of the related
datum.
   There might be, for example, some undetected systematic \cite{syst}
contribution to the reported result of the experiment measuring this
mass.
   We then reread the experimental paper with extra care for possible
uncertainties, and often ask the authors for further information.
   This often leads to corrections.  \\
   In the case where a mass determined from {\sc only one} experiment
(or from same experiments) deviate severely from the smooth surface,
replacement by an estimated value would give a more usefull information
to the user of the tables and would prevent obscuring the plots for the
observation of the mass surface.
   Fig.\,\ref{fig:s2n} for one of the derivatives of the mass surface
(cf.~Section~\ref{sect:deriv}) is taken from {\sc Ame'93} and shows how
replacements of a few such data by estimated values, can repair the
surface of masses in a region, not so well known, characterized by
important irregularities.
   Presently, only the most striking cases, not all
irregularities, have been replaced by estimates: typically those that
obscure plots like in Fig.\,\ref{fig:s2n}.

   \begin{figure}[htb]   
   \begin{center}
   \rotatebox{-90}{\includegraphics[bb=93 54 563 750,clip,height=15 cm]{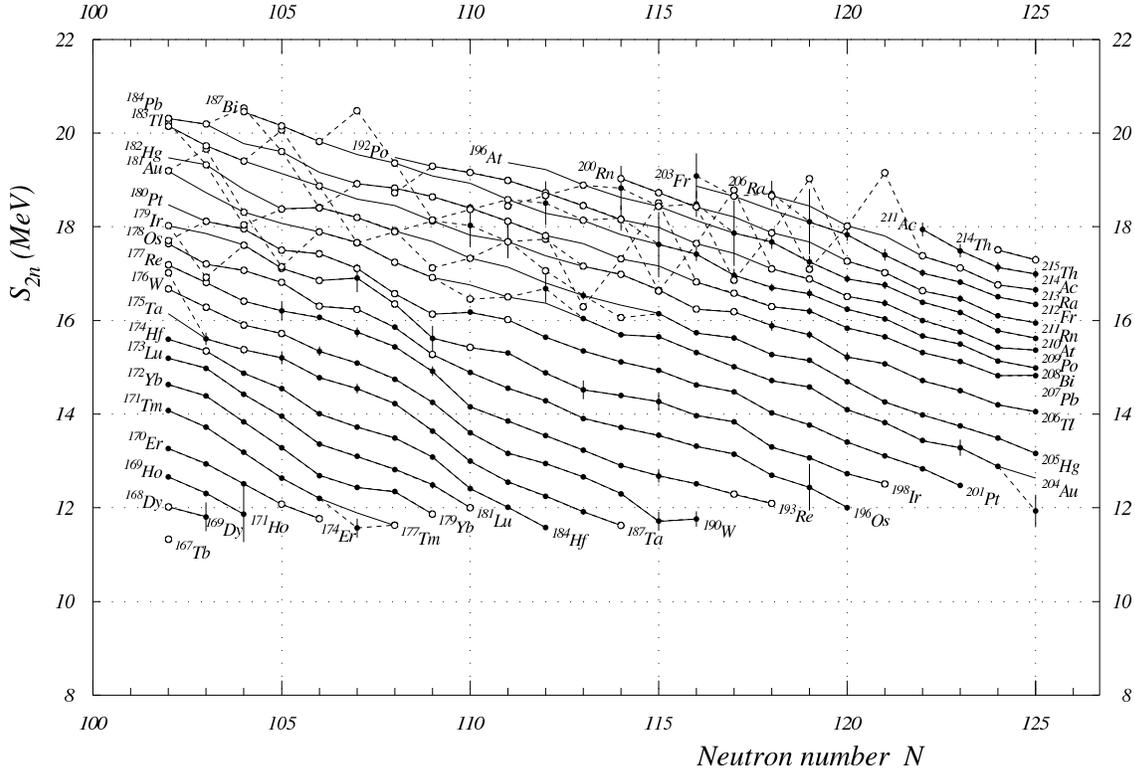}}
   \caption[]{\footnotesize
   Two-neutron separation energies as a function of $N$ (from
{\sc Ame'93}, p.~166).
   Solid points and error bars represent experimental values, open
circles represent masses estimated from ``trends in systematics".
   Replacing some of the experimental data by values estimated from
these trends, changes the mass surface from the dotted to the full
lines.
   The use of a `derivative' function adds to the confusion of the
dotted lines, since two points are changed if one mass is displaced.
   Moreover, in this region there are many $\alpha$ links resulting in
large propagation of errors.
   }\label{fig:s2n}
   \end{center}
   \end{figure}

   \item
   {\em Conflicts among data} - There are cases where some experimental data on the mass of a
particular nuclide disagree among each other and no particular reason
for rejecting one or some of them could be found from studying the
involved papers.
   In such cases, the measure of agreement with the just mentioned
regularity can be used by the evaluators for selecting which of the
conflicting data will be accepted and used in the evaluation.

   \item
   {\em Estimates} - Finally, drawing the mass surface allows to derive estimates for the
still unknown masses, either from interpolations or from short
extrapolations, as can be seen in Fig.\,\ref{fig:duf}.
   \begin{figure}[htb]   
   \begin{center}
   \rotatebox{-90}{\includegraphics[height=15cm]{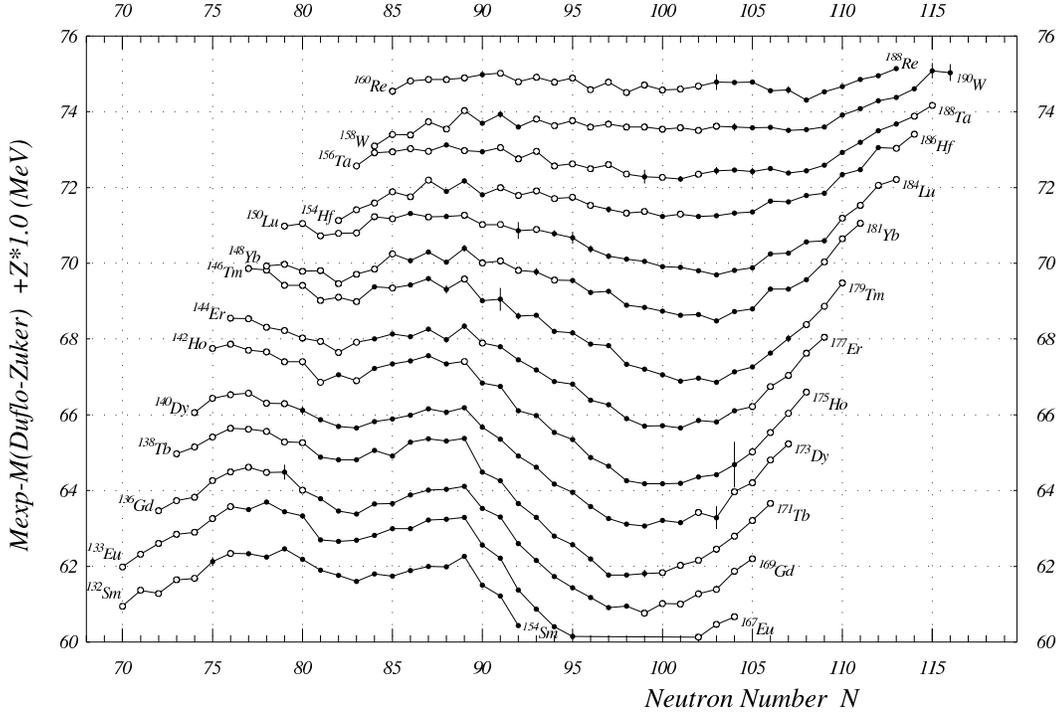}}
   \caption[]{\footnotesize
   Differences, in the rare-earth region, between the masses and the
values predicted by the model of Duflo and Zuker \protect\cite{95Duflo}.
   Open circles represent values estimated from systematic trends;
points are for experimental values.
   }\label{fig:duf}
   \end{center}
   \end{figure}

   \end{enumerate}

   \subsection{Extrapolations}

   In the case of extrapolation however, the error in the estimated mass
will increase with the distance of extrapolation.
   These errors are obtained by considering several graphs of
systematics with a guess on how much the estimated mass may change
without the extrapolated surface looking too much distorted.
   This recipe is unavoidably subjective, but has proven to be efficient
through the agreement of these estimates with newly measured masses in
the great majority of cases.

   It would be desirable to give estimates for all unknown nuclides that
are within reach of the present accelerator and mass separator
technologies.
   But, in fact, the {\sc Ame} only estimates values for all nuclides
for which at least one piece of experimental information is available
(e.g. identification or half-life measurement or proof of instability
towards proton or neutron emission).
   In addition, the evaluators want to achieve continuity in $N$, in
$Z$, in $A$ and in $N-Z$ of the set of nuclides for which mass values
are estimated.
   This set is therefore the same as the one defined for {\sc Nubase}
\cite{Nubase03}.

   To be complete, it should be said that {\sc regularity} is not the
only property used to make estimates: all available experimental
information is taken into account.
   The limits for the methods based on {\sc regularity} appear rapidly
when going down to low mass numbers where nuclear structures appear and
disappear on very short ranges, not to mention vanishing magic numbers.

   Neutron-rich masses could be constrained by our knowledge of
nuclides being stable or unstable relative to neutron emission: e.g.
stability against neutron emission implies a positive neutron separation
energie.
   This property, however, is usefull only for light species, where the
neutron drip-line can be reached.
   Similarly for proton-rich nuclides, but here one has to be carefull
and take into account the Coulomb barrier which may hinder proton emission
of a slightly p-unbound nucleus.
   Light proton-rich masses could be derived from the masses of their
mirror companions, or from the {\sc Imme} (Isobaric Multiplet Mass Equation)
which showed, up to now, to be fairly well verified.
   New developments in these directions are in progress.

   \subsection{Scrutinizing the surface of masses}

   Direct representation of the mass surface is not convenient since the
binding energy varies very rapidly with $N$ and $Z$.
   Splitting in four sheets, as mentioned above, complicates even more
such a representation.
   There are two ways to still be able to observe with some precision
the surface of masses: one of them uses the {\em derivatives} of this
surface, the other is obtained by {\em subtracting a simple function} of
$N$ and $Z$ from the masses.

   They are both described below and I will end this section with a
description of the interactive computer program that visualizes all
these functions to allow easier derivation of the estimated values.

   \subsubsection{The derivatives of the mass surface} \label{sect:deriv}

   By {\em derivative} of the mass surface we mean a specified
difference between the masses of two nearby nuclei.
   These functions are also smooth and have the advantage of displaying
much smaller variations.
   For a derivative specified in such a way that differences are between
nuclides in the same mass sheet, the near parallelism of these leads
to an (almost) unique surface for the derivative, allowing thus a single
display.
   Therefore, in order to illustrate the systematic trends of the
masses, four derivatives of this last type were traditionally chosen:
   \begin{enumerate}
   \item
   the two-neutron separation energies versus $N$, with lines connecting
the isotopes of a given element, as in Fig.\,\ref{fig:s2n};
   \item
   the two-proton separation energies versus $Z$, with lines connecting
the isotones (the same number of neutrons);
   \item
   the $\alpha$-decay energies versus $N$, with lines connecting the
isotopes of a given element; and
   \item
   the double $\beta$-decay energies versus $A$, with lines connecting
the isotopes and the isotones.
 \end{enumerate}
   These four derivatives are given in the printed version of the {\sc Ame2003},
Part~II, Figs.~1--36 \cite{Ame03a}.

   However, from the way these four derivatives are built, they give
only information within one of the four sheets of the mass surface (e-e,
e-o, o-e or e-e; e-o standing for even $N$ and odd $Z$).
   When observing the mass surface, an increased or decreased spacing of
the sheets cannot be observed.
   Also, when estimating unknown masses, divergences of the four sheets
could be unduly created, which is unacceptable.

   Fortunately, other various representations are possible (e.g.
separately for odd and even nuclei: one-neutron separation energies
versus $N$, one-proton separation energy versus $Z$, $\beta$-decay
energy versus $A$, \ldots).
   Such graphs have been prepared and can be obtained from the {\sc Amdc}
web distribution \cite{amdcgr}.

   The method of `derivatives' suffers from involving two masses for
each point to be drawn, which means that if one mass is moved then two
points are changed in opposite direction, causing confusion in our
drawings Fig.\,\ref{fig:s2n}.

   \subsubsection{Subtracting a simple function} \label{sect:func}

   Since the mass surface is smooth, one can try to define a function of
$N$ and $Z$ as simple as possible and not too far from the real surface
of masses.
   The difference between the mass surface and this function, while
displaying reliably the structure of the former, will vary much less
rapidly, improving thus its observation.

   \paragraph{       Subtracting results from a model} \label{sect:submod}

   Practically, we use the results of the calculation of one of the
modern models.
   However, we can use here only those models that provide masses
specifically for the spherical part, forcing the nucleus to be
un-deformed.
   The reason is that the models generally describe quite well the
shell and sub-shell closures, and to some extent the pairing energies,
but not the locations of deformation.
   If the theoretical deformations were included and not located at
exactly the same position as given by the experimental masses, the mass
difference surface would show two dislocations for each shape
transition.
   Interpretation of the resulting surface would then be very difficult.

   My two choices are the ``New Semi-Empirical Shell Correction to
the Droplet Model (Gross Theory of Nuclear Magics)" by Groote, Hilf and
Takahashi \cite{76Groote};
   and the ``Microscopic Mass Formulas" of Duflo and Zuker
\cite{95Duflo}, which has been illustrated above (Fig.\,\ref{fig:duf}).
   In {\sc Ame2003}, we made extensive use of such differences with models.
   The plots we have prepared were not published, they can be retrieved
from the {\sc Amdc} \cite{amdcgr}.

   The difference of mass surfaces shown in Fig.\,\ref{fig:duf} is
instructive:
   \begin{enumerate}
   \item
   the lines for the isotopic series cross the $N$=82 shell closure with
almost no disruption, showing thus how well shell closures are described
by the model;
   \item
   the well-known onset of deformation in the rare-earth at $N$=90
appears very clearly here as a deep large bowl, since deformation is not
used in this calculation.
   The contour of this deformation region is neat.
   The depth, i.e. the amount of energy gained due to deformation,
compared to ideal spherical nuclides, can be estimated; and
   \item
   Fig.\,\ref{fig:duf} shows also how the amplitude of deformation
decreases with increasing $Z$ and seems to vanish when approaching
Rhenium ($Z$=75).
   \end{enumerate}

   \paragraph{       Subtracting a Bethe and Weizs\"acker formula} \label{sect:subbw}

   Since 1975 \cite{75Thib}, and now with more and more evidence, it
appears that the {\em `magic'} numbers that were though to occur at
fixed numbers of protons or neutrons, are not really so.
   They might disappear when going away from the valley of stability, or
appear at new locations.
   Where such migrations occur, most models are much in trouble, and the
reasoning we made above for not including deformation in the models,
now applies also to `magic' numbers.
   Excluding shell and sub-shell closures, we are then driven directly
back to the pioneering work of Weizs\"acker \cite{35Weizs}.
   Recent works \cite{03Fletcher,04vanisa} on the modern fits of the
original Bethe and Weizs\"acker's formula seems promising in this
respect.

   The concept of the liquid drop mass formula was defined by
Weizs\"acker in 1935 \cite{35Weizs} and fine-tuned by Bethe and Bacher
\cite{36Bethe} in 1936.
   The binding energy of the nucleus comprises only a volume energy
term, a surface one, an asymmetry term, and the Coulomb energy
contribution for the repulsion amongst protons.
   The {\em total} mass is thus:
   \begin{equation}\label{equ:bethe}
   \Ma(N,Z) = N\Ma_{\rm n} + Z\Ma_{\rm H} - \alpha A +
   \beta \frac{(N-Z)^2}{A} + \gamma A^{2\over3} +
   \frac{3}{5} \frac{e^2Z^2}{r_0A^{1\over3}}
   \end{equation}
   where $A=N+Z$, is the atomic weight, $r_0A^{1/3}$ the nuclear radius,
$\Ma_{\rm n}$ and $\Ma_{\rm H}$ the masses of the neutron and of the
hydrogen atom.
   The constants $\alpha$, $\beta$, $\gamma$ and $r_0$ were determined
empirically by Bethe and Bacher:
   $\alpha=13.86$~MeV,
   $\beta=19.5$~MeV,
   $\gamma=13.2$~MeV and
   $r_0=1.48\cdot 10^{-15}$~m
   (then $\frac{3}{5} e^2/r_0=0.58$~MeV).
   The formula of Eq.\,(\ref{equ:bethe}) is unchanged if $\Ma(N,Z)$,
$\Ma_{\rm n}$ and $\Ma_{\rm H}$ are replaced by their respective mass
excesses (at that time they were called {\em mass defects}).
   When using the {\em constants} given above one should be aware that
when Bethe fixed them, he used for the mass excesses of the neutron and
hydrogen atom respectively 7.8~MeV and 7.44~MeV in the $^{16}$O standard,
with a value of 930~MeV for the atomic mass unit.

   If we subtract Eq.\,(\ref{equ:bethe}) from all masses we are left
with values that vary much less rapidly than the masses themselves,
while still showing all the structures.
   However, the splitting in four sheets will still make the image
fuzzy.
   One can then add to the right hand side of the formula of Bethe
(\ref{equ:bethe}) a commonly used pairing term
$\Delta_{pp}=\Delta_{nn}=-12/\sqrt{A}$~MeV and no $\Delta_{np}$
(Fig.\,\ref{fig:pairing}), which is sufficient for our purpose.
   (For those interested, there is a more refined study of the
variations of the pairing energies that has been made by Jensen, Hansen
and Jonson \cite{84Jensen}).

   \subsection{Far extrapolations} \label{sect:subfar}

   When exploiting these observations one can make extrapolations for
masses very far from stability.
   This has been done already \cite{98Borcea}, but with a further
refinement of this method obtained by constructing an {\em idealized}
surface of masses (or {\em mass-geoid}) \cite{95Borcea}, which is the
best possible function to be subtracted from the mass surface.
   In Ref.\,\cite{98Borcea}, a local {\em mass-geoid} was built as a
cubic function of $N$ and $Z$ in a region limited by magic numbers for
both $N$ and $Z$, fitted to only the purely spherical nuclides and
keeping only the very reliable experimental masses.
   Then the shape of the bowl (for deformation) was reconstructed `by
hand', starting from the known non-spherical experimental masses.
   It was found that the maximum amplitude of deformation amounts to
5~MeV, is located at $^{168}$Dy, and that the region of deformation
extends from $N$=90 to $N$=114 and from $Z$=55 to $Z$=77, which is
roughly in agreement with what is indicated by Fig.\,\ref{fig:duf}.


   \begin{figure}   
   \begin{center}
   \includegraphics[width=14 cm]{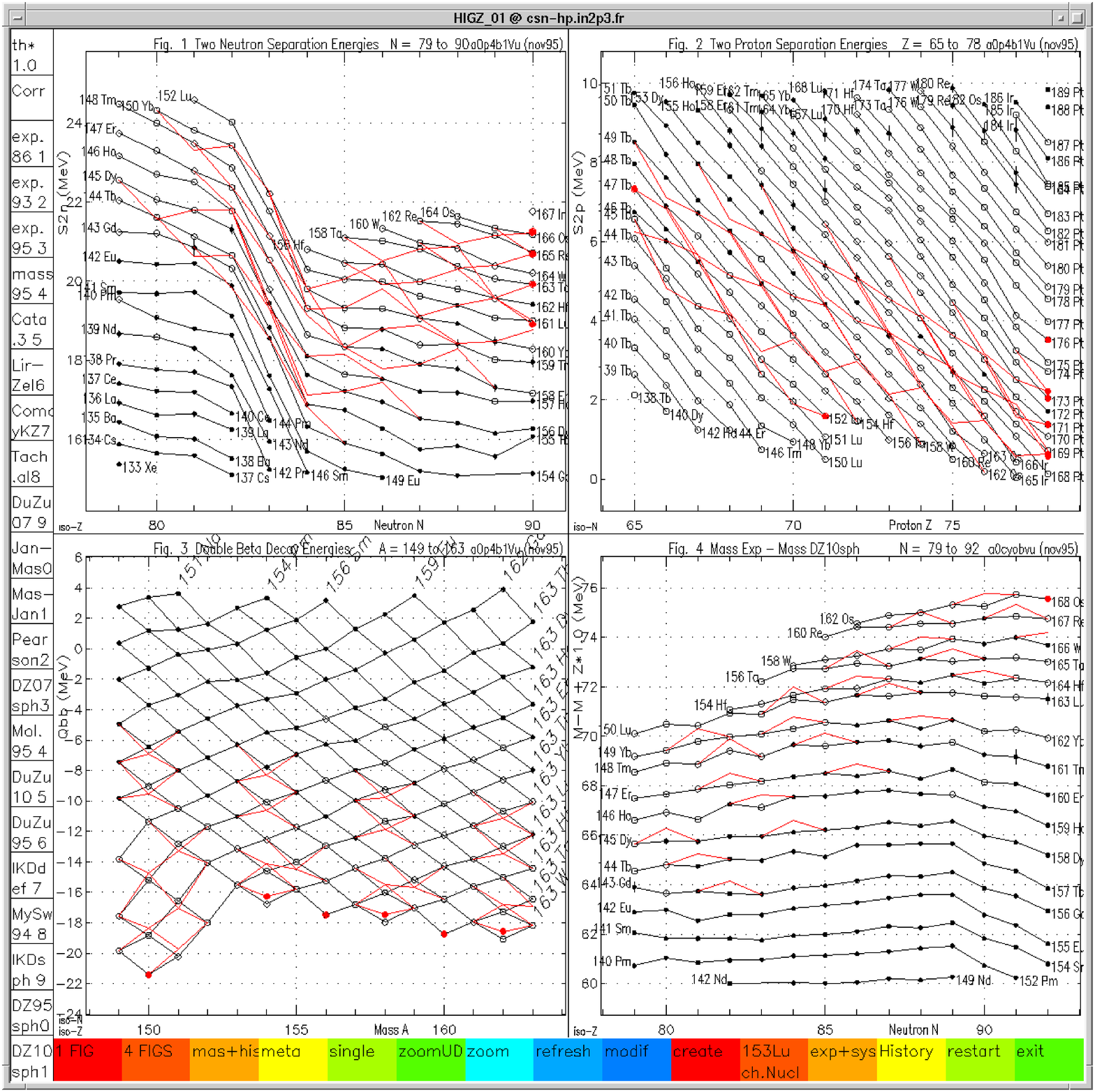}
   \caption[]{\footnotesize
   A screen image of {\sc Desint}, the interactive graphical display of
four cuts in the surface of masses around $^{146}$Gd.
   The four quadrants display respectively $S_{2n}(N)$, $S_{2p}(Z)$,
$Q_{2\beta}(A)$ and $(M_{\rm exp}-M_{\rm Duflo-Zuker})(N)$
\protect\cite{95Duflo}.
   The lines in black connect nuclides with same $Z$, $N$, ($Z$ and $N$)
and $Z$ respectively.
   The boxes at left and bottom serve for various interactive commands.
   The $N$=82 shell closure is clearly seen in quadrant 1 and in the
lower left corner of quadrant 3.
   The lines in red illustrate the many consequences of an increase of
the mass of $^{146}$Gd by 500~keV.
   }\label{fig:desint}
   \end{center}
   \end{figure}

   \subsection{Manipulating the mass surface}

   In order to make estimates of unknown masses or to test changes on
measured ones, one needs to visualize different graphs, either from the
`derivatives' type or from the `difference' type.
   On these graphs, one needs to add (or move) the relevant mass and
determine how much freedom is left in setting a value for this mass.

   Things are still more complicated, particularly for changes on
measured masses, since other masses could depend on the modified one,
usually through secondary data.
   Then one mass change may give on one graph several connected changes.

   Another difficulty is that a mass modification (or a mass creation)
may look acceptable on one graph, but may appear unacceptable on another
graph.
   One should therefore be able to watch several graphs at the same
time.

   A supplementary difficulty may appear in some types of graphs where
two tendencies may alternate, following the parity of the proton or of
the neutron numbers.
   One may then wish, at least for better comfort, to visualize only
one of these two parities.

   All this has become possible with the `interactive graphical tool',
called {\sc Desint} (from the French: `dessin interactif') written by
C.~Borcea \cite{93Borcea} and illustrated in Fig.\,\ref{fig:desint}.
   Any of the `derivatives' or of the `differences' can be displayed in
any of the four quadrants of Fig.\,\ref{fig:desint}, or alone and
enlarged.
   Any of these functions can be plotted against any of the parameters
$N$, $Z$, $A$, $N-Z$, and $2Z-N$; and connect iso-lines in any single or
double parameters of the same list (e.g., in the third view of
Fig.\,\ref{fig:desint}, iso-lines are drawn for $Z$ {\sc and} for $N$).
   Zooming in and out to any level and moving along the two coordinates
are possible independently for each quadrant.
   Finally, and more importantly, any change appears, in a different
color, with all its consequences and in all four graphs at the same
time.
   As an example and only for the purpose of illustration, a change of
+500~keV has been applied, in Fig.\,\ref{fig:desint}, to $^{146}$Gd in
quadrant number four; all modifications in all graphs appear in red.

   \section{The Tables}

   In December 2003, we succeeded in having published {\sc together}
the ``Atomic Mass Evaluation" {\sc Ame} \cite{Ame03,Ame03a} and the
{\sc Nubase} evaluation \cite{Nubase03}, which have the same ``horizontal"
structure and basic interconnections at the level of isomers.

   After the 1993 tables ({\sc Ame'93}) it was projected to have updated
evaluations performed regularly (every two years) and published in paper
only partly, while all files should still be distributed on the Web.
   Effectively, an update {\sc Ame'95} \cite{Ame95} appeared two years later.
   Lack of time to evaluate the stream of new quite important data, and
also the necessity to create the {\sc Nubase} evaluation (see below),
prevented the intended further updates of the {\sc Ame}.
   The {\sc Nubase} evaluation was thus published for the first time in
September 1997 \cite{Nubase97}, but in order to have consistency between
the two tables, it was decided then that the masses in {\sc Nubase'97}
should be exactly those from {\sc Ame'95}.
   A certain stabilization, that seems to be reached now, encouraged us
to publish in 2003 a new full evaluation of masses, together with the
new version of {\sc Nubase}.
   This time, the {\sc Ame2003} and {\sc Nubase2003} are completely
`synchronized'.

   Full content of the two evaluations is accessible on-line at the
web site of the Atomic Mass Data Center ({\sc Amdc}) \cite{amdc} through
the {\em World Wide Web}.
   One will find at the {\sc Amdc}, not only the published material, but
also extra figures, tables, documentation, and more specially the
{\sc Ascii} files for the {\sc Ame2003} and the {\sc Nubase2003} tables,
for use with computer programs.

   The contents of {\sc Nubase} can be displayed
   with a PC-program called ``{\sc Nucleus}" \cite{95Potet},
   and also
   by a Java program {\sc jvNubase} \cite{97Durand} through the
{\em World Wide Web},
   both distributed by the {\sc Amdc}.

   \section{Conclusion}

   Deriving a mass value for a nuclide from one or several experiments
is in most cases not easy.
   Some mathematical tools (the least-squares method) and computer tools
(interactive graphical display), and especially the evaluator's judgment
are essential ingredients to reach the best possible recommended values
for the masses.

   Unknown masses close to the last known ones can be predicted
from the extension of the mass surface.
   However, for the ones further out, more particularly those which are
essential in many astrophysical problems, like the nucleosynthesis
r-process, values for the masses can only be derived from some of the
available models.
   Unfortunately, the latter exhibit very large divergences among them
when departing from the narrow region of known masses, reaching up to
tens of MeV's in the regions of the r-process paths.
   Therefore, one of the many motivations for the best possible
evaluation of masses is to get the best set of mass values on which
models may adjust their parameters and better predict masses further away.

   \section*{Acknowledgements }

   I would like to thank Aaldert~H.~Wapstra with whom I have been
working since 1981.
   The material used in this lecture is also his material.
   He was the one who established in the early fifties the {\sc Ame} in
its modern shape as we know now.
   Aaldert~H.~Wapstra has always been very accurate, very careful and
hard working in his analysis in both the {\sc Ame} and the {\sc Nubase}
evaluations.
   During these 23 years I have learned and still learn a lot from his
methods.
   I wish also to thank my close collaborators:
   Jean Blachot who triggered in 1993 the {\sc Nubase} collaboration,
   Olivier Bersillon,
   Catherine Thibault,
   and Catalin Borcea who built the computer programs for mass
extrapolation, and worked hard at the understanding, the definition and
the construction of a {\em mass-geoid}.


\begin{thebibliography}{00}
\bibitem{syst}                     
     Systematic errors are those due to instrumental drifts or
     instrumental fluctuations, that are beyond control and are
     not accounted for in the error budget.
     They might show up in the calibration process, or when the
     measurement is repeated under different experimental conditions.
     The experimentalist adds then quadratically a systematic error to
     the statistical and the calibration ones, in such a way as to have
     consistency of his data.
     If not completely accounted for or not seen in that experiment,
     they can still be observed by the mass evaluators when considering
     the mass adjustment as a whole.
\end{thebibliography}
\end{document}